\documentclass[hidelinks,a4paper,10pt,table]{article}
\usepackage{amsmath}
\usepackage{amsthm}
\usepackage{amssymb}
\usepackage{adjustbox} % rotate table 1
\usepackage{algorithmic}
\usepackage{color}
\usepackage{caption,subcaption}
\usepackage{graphicx}
\usepackage{hyperref} %% sections
\usepackage{mathrsfs}
\usepackage[margin=0.8in]{geometry}
\usepackage{tikz}
\usepackage{wrapfig}
\usepackage{rotating}
\usepackage{booktabs}
\usepackage{array}
\usepackage{multirow, makecell}
\usepackage[round]{natbib}
\usepackage{bm} %% bold
\usepackage{bigints} %% for big integral
\usepackage{float} %% for [H]
\usepackage{bbm} %% for \mathbbm{1}
\usepackage[ruled,vlined,linesnumbered]{algorithm2e} % for algorithm
\usepackage{setspace}
\usepackage{authblk}

\DeclareMathOperator*{\argmax}{arg\,max}

\DeclareMathOperator{\Var}{Var}
\DeclareMathOperator{\cov}{Cov}

\DeclareMathOperator{\logit}{logit}

\newtheorem{assumption}{Assumption}
\theoremstyle{definition}

\title{Bayesian integration G-formula for platform SMART designs allowing for adding new treatments}
\author[1]{Xinru Wang}
\author[1]{Meghna Bose}
\author[1,2,3]{Bibhas Chakraborty} 
\author[4,*]{Robert Mahar}

\affil[1]{Centre for Biomedical Data Science, Duke-NUS Medical School, Singapore}
\affil[2]{Department of Statistics and Data Science, National University of Singapore, Singapore}
\affil[3]{Department of Biostatistics and Bioinformatics, Duke University, North Carolina, USA}
\affil[4]{Melbourne School of Population and Global Health, University of Melbourne, Victoria, Australia}
\affil[*]{\it Correspondence: Robert Mahar (robert.mahar@unimelb.edu.au)}

\date{}
\begin{document}
\onehalfspacing
\maketitle

\begin{abstract}
Dynamic treatment regimes (DTRs) are sequences of decision rules to guide treatment assignments in response to a patient's evolving, time-varying disease status. Sequential multiple assignment randomized trials (SMARTs) are considered the gold standard experimental design for evaluating DTRs. However, SMARTs often require more time to complete compared with a single stage RCT and new candidate treatments may become available or feasible during the trial. Platform trials are an adaptive trial design that allow new treatments to be added to the ongoing study according to a prespecified master protocol. In this paper, we introduce a novel platform SMART that integrates features from both platform trials and SMARTs, allowing new treatments to be added during the trial. Additionally, we propose the Bayesian integration G-formula (BIG) estimators for platform SMARTs to account for non-concurrent treatment comparisons. Extensive simulations are conducted to evaluate the performance of different BIG estimators against benchmark methods. We demonstrate the proposed BIG estimators based on the \textit{S. aureus} Network Adaptive Platform (SNAP) trial. 
\end{abstract}

\vspace{5pt}

\section{Introduction}
In the era of big data and digital innovation, healthcare has been revolutionized, transitioning from the {\it one-size-fits-all} standard to a personalized approach catering to each patient’s individual needs. Under this framework, the treatments recommended to patients are tailored according to their individual characteristics (e.g., age, gender, social status, lifestyle) or the evolving dynamics (like response to initial treatment or disease progression over time), often involving several stages of randomization and re-randomization. It ensures that the right drug with the right dosage is offered to the right patients at the right time. In the past, heterogeneous treatment effects among patients have been observed in the practical clinical management of chronic and recurring illnesses. For example, one of the most common diseases, cancer, affects individuals differently, and physicians often make repeated decisions about the best way to continue a patient's treatment based on their disease progression over time \citep[see][]{kidwell2014smart}. These kinds of targeted therapies are also common in the case of mental health disorders like Attention Deficit Hyperactive Disorder (ADHD), where low-dose medications initially offered may be escalated to higher doses or augmented with behavioral interventions depending on the patient's individual response \citep[see][]{nahum2012experimental}. Treatments administered in multiple stages, with dosage, type, or intensity adjusted based on the patients' disease condition over time, obviously lead to improved treatment effectiveness and better management of side effects. However, finding the ideal treatment sequence for a patient is the primary interest, mathematically formulated as the problem of finding the optimal dynamic treatment regimes (DTRs). DTRs are sequences of decision rules that formalize sequential treatment strategies that are based on the patient's time-varying characteristics \citep{chakraborty2013statistical}. Estimating optimal DTRs based on data from either sequential multiple assignment randomized trials (SMARTs) or observational studies requires different and often complicated methods depending on the DTRs in question, the outcome type, and the research questions of interest \citep{chakraborty2013statistical, mahar2021scoping, chakraborty2016estimating, schulte2014q, zhang2018interpretable, wang2023sequential}. 

SMARTs are experimental designs with multiple randomization stages and are considered the gold-standard design for estimating optimal DTRs \citep{murphy2005experimental, wang2023sequential}. By randomizing participants more than once, SMARTs provide participants a second opportunity to receive a better treatment in the trial, while also providing unbiased information for constructing optimal DTRs. However, because there are multiple randomization stages, SMARTs often require more time to complete compared with single-stage RCTs \citep{wu2023interim}. Extended completion times can pose challenges, particularly if the new treatments that were not initially suitable for inclusion in a SMART become available during the trial because of, for example, regulatory approval \citep{cohen2015adding}. 

Certain types of adaptive trials, such as platform trials, allow new treatments to be incorporated into an ongoing study following a master protocol, rather than conducting a separate study \citep{berry2015platform} to compare the newly available treatments with the control. By answering new research questions using existing trial infrastructure, platform trials can conserve trial resources, minimize participant burden, and streamline trial procedures. Besides, it is reasonable to offer participants the opportunity to receive newly approved drugs instead of continuing with the treatment options in the original trial. Many large-scale platform trials are currently underway across diverse disease areas \citep{griessbach2024characteristics}, for example, oncology \citep{sydes2012flexible}, infectious diseases \citep{recovery2022casirivimab, tong2022staphylococcus}, and neurology \citep{wong2022motor}, with many more under development. 

Combining features of platform trials with SMARTs presents an opportunity for improvement. Indeed, there are multiple platform trials underway that have underlying SMART characteristics such as treatment randomization that depends on intermediate outcomes \citep{tong2022staphylococcus, mahar2023blueprint, wainright_finding_2023}, although the primary analysis is not of constructing optimal DTRs, but rather stage-specific treatment effects. One of the examples is the \textit{S. aureus} Network Adaptive Platform (SNAP) trial, which aims to evaluate the effects of multiple treatments for different subgroups of patients with \textit{S. aureus} bloodstream infection \citep{tong2022staphylococcus}. In the SNAP trial, participants were first randomized among different antibiotic intervention in both a `backbone' and an `adjunctive' treatment domain. Those who responded to the initial treatment (at either 7 or 14 days post enrolment into the trial) were randomized to either the standard care or switch to oral antibiotics; while those who did not show satisfactory response continued with the standard care. In the SNAP trial, additional treatments may be introduced throughout the trial, while ineffective treatments can be removed based on interim analysis findings. 

While adding new treatments to a trial can save time and resources, it also introduces several statistical challenges \citep{marschner2023transparent, bofill2023use, lee2020including}. An ongoing debate revolves around whether to use data from the nonconcurrent group to enhance the trial's efficiency or to solely rely on data from concurrently randomized cohorts. Changes over time in the available treatment set, randomization probabilities, underlying population, or the standard care, can confound estimates of treatment effect, potentially leading to biases. Different approaches have been proposed to mitigate this confounding, either from randomization perspectives \citep{ventz2018adding, viele2023allocation}, or analytical perspectives including model-based approaches \citep{bofill2023use, viele2023allocation,saville2022bayesian}, network meta-analytical approaches \citep{marschner2022analysis}, 
Bayesian approaches that use the form of dynamic borrowing mechanism -- dynamic in the sense that it adapts the amount of borrowing depending on the amount of conflict between nonconcurrent controls and the current ones \citep{hobbs2011hierarchical, schmidli2014robust, krotka2023ncc}. All of these approaches balance a tradeoff between statistical efficiency and possible bias, with the most appropriate method depending on the research context. 

Despite the increasing popularity of platform trials, to the best of our knowledge, currently there are no studies that focus on analytical approaches for SMARTs that allow for adding new treatments (hereafter referred to as `platform SMARTs'). In this paper, we propose novel Bayesian integration G-formula (BIG) estimators that combine DTR data across different stages of a platform SMART for comparing embedded DTRs. We conduct a simulation study to evaluate the performance of BIG estimators with different priors against the benchmark approaches that use either crude estimates obtained from data pooled over \textit{pre-adaptation} and \textit{post-adaptation} stages or obtained using only \textit{post-adaptation} data from a platform SMART. In addition, we demonstrate the practical value of the proposed method using the SNAP trial as a case study. . 

The remainder of this paper proceeds as follows. Section~\ref{section2} provides the basic set-up and notations for two types of two-stage SMARTs: one without the introduction of new treatments and one with an additional treatment at the first stage. Section~\ref{section3} introduces the BIG estimators with various prior distributions for estimating DTR means using data from platform SMARTs. In section~\ref{section4}, we evaluate the operating characteristics using each method in terms of the probability of identifying the true optimal DTR, bias, variance, mean squared error (MSE), and the coverage rate (CR) of the confidence interval for the estimated difference in DTR means. We illustrate our proposed method using the SNAP trial in section~\ref{section5} and conclude this paper with a discussion and propose potential avenues to extend our contributions in Section~\ref{section6}.

\section{Notations in two-stage SMART and platform SMARTs}
\label{section2}

To better articulate our approach, we start with a fixed two-stage SMART and then proceed to a platform SMART with an additional initial treatment added during the trial. 

\subsection{Fixed two-stage SMART}

\begin{figure}
\centering
\includegraphics[width=0.4\textwidth]{"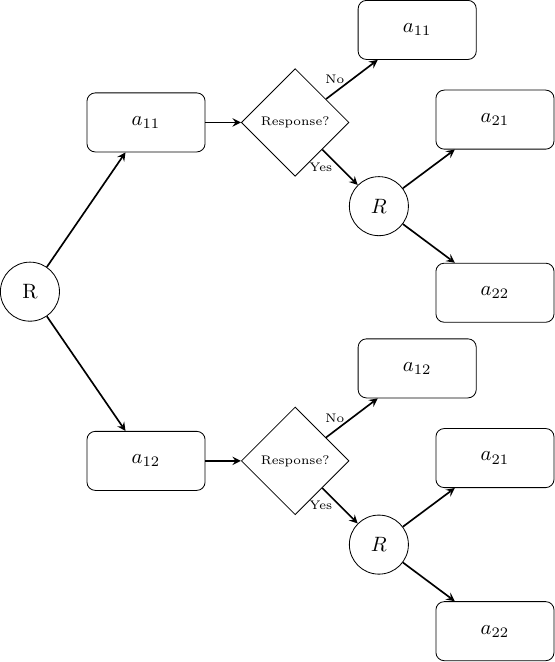"}
\caption{A conventional two-stage SMART. ``$R$'' denotes randomization.}
\label{figure:conventional_SMART}
\end{figure}

We restrict our attention to a simplified two-stage SMART design based on the SNAP trial \citep{tong2022staphylococcus} shown in \autoref{figure:conventional_SMART}. Participants are initially randomized to one of two treatments, denoted as $a_{1j} \in \mathcal{A}_1$, where $\mathcal{A}_1$ is the treatment domain in the first stage and $a_{1j}$ is the $j$-th treatment in $\mathcal{A}_1$. The first stage includes a total of $J=2$ treatments. Let $R_i \in \{0,1 \}$ denote the observed intermediate response status for the $i$-th participant where $R_i=1$ and $R_i=0$ indicate response and non-response to the initial treatment, respectively. In the second stage, non-responders continue with the initial treatment, while responders are randomized among two treatments denoted as $a_{2k} \in \mathcal{A}_2$, where $a_{2k}$ and $\mathcal{A}_2$ are defined similarly as those in stage 1. The observed data for the $i$-th participant can be denoted as $\{\bm{X}_{1i}, A_{1i}, \bm{X}_{2i}, A_{2i},  Y_i\}$, where $\bm{X}_{1i}$ and $\bm{X}_{2i}$ represent the vector of baseline and potential tailoring covariates and $\bm{X}_{2i}$ includes the response status $R_i$. Hereafter, unless otherwise stated, we use uppercase letters to represent random variables, lowercase letters to represent the corresponding observed values, and bold letters to represent vectors when there is no confusion. 

This SMART has four embedded DTRs denoted as $d_{jk} = a_{1j} a_{1j}^{1-R}a^{R}_{2k}$, which can be interpreted as `first treat patients with $a_{1j}$, continue treatment $a_{1j}$ if they do not respond and switch to $a_{2k}$ if they respond'. Let $Y_i$ denote the final outcome for the $i$-th participant, and without loss of generality, we assume that $Y_i$ is a continuous variable and that higher values of $Y_i$ are preferable. The overall objective is to select the optimal DTR based on the expected mean outcomes under each embedded DTR. 

Let $R_{i}(a_{1j}) \in \{0,1\}$ denote the counterfactual response status for the $i$-th participant under the initial treatment $a_{1j}$. Let $Y_i(a_{1j}a_{1j})$ be the counterfactual outcome for the $i$-th participant who receives treatment $a_{1j}$, does not respond, and continues with $a_{1j}$ and let $Y_i(a_{1j}a_{2k})$ be the counterfactual outcome for the $i$-th participant who receives treatment $a_{1j}$, responds, and switches to $a_{2k}$. The counterfactual outcome for the $i$-th participant who follows the DTR $d_{jk}$ is defined as
\begin{equation}
    Y_i(d_{jk}) = R_{i}(a_{1j}) Y_i(a_{1j}a_{2k}) + (1-R_{i}(a_{1j})) Y_i(a_{1j}a_{1j}).
\end{equation}
Throughout the paper, we assume the general causal assumptions under the Neyman-Rubin causal inference framework \citep{rubin1974estimating}. Let $R_{i}(a_{1j}) \in \{0,1\}$ denote the counterfactual response status for the $i$-th participant under the initial treatment $a_{1j}$. Let $Y_i(a_{1j}a_{1j})$ be the counterfactual outcome for the $i$-th participant who receives treatment $a_{1j}$, does not respond, and continues with $a_{1j}$ and let $Y_i(a_{1j}a_{2k})$ be the counterfactual outcome for the $i$-th participant who receives treatment $a_{1j}$, responds, and switches to $a_{2k}$. The counterfactual outcome for the $i$-th participant who follows the DTR $d_{jk}$ is defined as
\begin{equation}
    Y_i(d_{jk}) = R_{i}(a_{1j}) Y_i(a_{1j}a_{2k}) + (1-R_{i}(a_{1j})) Y_i(a_{1j}a_{1j}).
\end{equation}
We assume the following: 1) consistency, i.e., the observed variables correspond to the potential variables under the treatment received; 2) exchangeability, i.e., the randomization probability is independent of the potential outcomes; and 3) positivity, i.e., the randomization probability is none zero values for each treatment options. Let $\pi_j$ denote the response rate for treatment $a_{1j}$, the expected outcome for the DTR $d_{jk}$ can then be denoted as 
\begin{equation}
    \mu_{jk} = E[Y_i(d_{jk})] = \pi_j\mu_{a_{1j}a_{2k}} + (1-\pi_j) \mu_{a_{1j}a_{1j}}, \quad j,k = 1,2,
\end{equation}
where $\mu_{a_{1j}a_{1j}}= E[Y(a_{1j}a_{1j})]$ and $\mu_{a_{1j}a_{2k}}= E[Y(a_{1j}a_{2k})]$ are the expected outcomes for those who follow treatment sequence $a_{1j}a_{1j}$ and those who follow treatment sequence $a_{1j}a_{2k}$, respectively. The optimal DTR is then defined as $d^{\ast} = \underset{d_{jk} \in \mathcal{D}}{\argmax \mu_{jk}}$, where $\mathcal{D}$ is the set of available DTRs.

Under the Neyman-Rubin causal inference framework \citep{rubin1974estimating}, the inverse probability weighting (IPW) estimator \citep{ogbagaber2016design} for the expected outcome under the DTR $d_{jk}$ is 
\begin{equation}
\label{mu_1}
    \hat{\mu}_{jk} = \frac{\sum_{i=1}^n w_i^{(jk)} Y_i}{\sum_{i=1}^n w_i^{(jk)}}, 
\end{equation}
where $w_i^{(jk)} = \left\{ \frac{(1-R_i)I_i(a_{1j})}{p_j} + \frac{R_i I_i(a_{1j})I_i(a_{2k})}{p_j q_k} \right\}$, $I_i(\cdot)=1$ if the participant receives the treatment in the indicator function, $I_i(\cdot)=0$ otherwise, $n$ is the sample size, and $p_j$ is the randomization probability for treatment $a_{1j}$ at the first stage and $q_k$ is the randomization probability for treatment $a_{2k}$ at the second stage. The estimated optimal DTR is thus $\hat{d}^{\ast} = \underset{d_{jk} \in \mathcal{D}}{\argmax \hat{\mu}_{jk}}$. The asymptotic variance of $\hat{\mu}_{jk}$ is 
\begin{equation}
\label{eq_var1}
\begin{aligned}
    \Var(\hat{\mu}_{jk})  =\frac{\sigma^2_{jk}}{n} &= \frac{1}{n} \left[ \frac{\pi_j}{p_j q_k} \left \{ \sigma^2_{a_{1j}a_{2k}} + (1-\pi_j)^2 (\mu_{a_{1j}a_{1j}} - \mu_{a_{1j}a_{2k}})^2 \right \}  \right.\\
    &\left. + \frac{(1-\pi_j)}{p_j}  \left \{ \sigma^2_{a_{1j}a_{1j}} + \pi_j^2 (\mu_{a_{1j}a_{1j}} - \mu_{a_{1j}a_{2k}})^2 \right \} \right],
\end{aligned} 
\end{equation}
where $\sigma^2_{a_{1j}a_{1j}}= \Var(Y_i(a_{1j}a_{1j}))$ and $\sigma^2_{a_{1j}a_{2k}} = \Var(Y_i(a_{1j}a_{2k}))$. The asymptotic covariance for the two DTRs $d_{jk}$ and $ d_{jk'}$ that share the same initial treatment is denoted as 
\begin{equation}
\label{eq_var2}
    \begin{aligned}
        \cov(\hat{\mu}_{jk}, \hat{\mu}_{jk'}) &= \frac{\sigma_{jk,jk'}}{n}\\& = \frac{(1-\pi_j )}{p_j n}\times \left\{ \sigma^2_{a_{1j}a_{1j}} + \pi_j^2(\mu_{a_{1j}a_{1j}}-\mu_{a_{1j}a_{2k}})  (\mu_{a_{1j}a_{1j}}-\mu_{a_{1j}a_{2k'}})\right\}, ~ k\neq k'. 
        \end{aligned}
\end{equation}
The covariance between DTRs that start with different initial treatments is zero, i.e., $\cov(\hat{\mu}_{jk}, \hat{\mu}_{j'k}) = \cov(\hat{\mu}_{jk}, \hat{\mu}_{j'k'}) = 0$ for $j \neq j'$.

The above variance and covariance can be estimated from SMART data either by plugging in the estimated parameters in Eq.~(\ref{eq_var1}) and Eq.~(\ref{eq_var2}) or by the following equations \citep{ko2012up, ogbagaber2016design}:
\begin{equation}
\label{eq_var3}
\begin{aligned}
    \widehat{\Var}(\hat{\mu}_{jk}) &= \frac{\hat{\sigma}_{jk}}{n} = \frac{1}{n^2} \sum_{i=1}^n \left\{ \left(\frac{(1-R_i)I_i(a_{1j})}{p_j} + \frac{R_i I_i(a_{1j})I_i(a_{2k})}{p_j q_k} \right) (Y_i-\hat{\mu}_{jk}) \right\}^2,
\end{aligned} 
\end{equation}
and
\begin{equation}
\label{eq_var4}
\begin{aligned}
    \widehat{\cov}(\hat{\mu}_{jk}, \hat{\mu}_{jk'}) = \frac{\hat{\sigma}_{jk,jk'}}{n} &= \frac{1}{n^2} \sum_{i=1}^n \left\{ \left(\frac{(1-R_i)I_i(a_{1j})}{p_j} + \frac{R_i I_i(a_{1j})I_i(a_{2k})}{p_j q_k} \right) (Y_i-\hat{\mu}_{jk}) \right. \\ 
    &\times \left. \left(\frac{(1-R_i)I_i(a_{1j})}{p_j} + \frac{R_i I_i(a_{1j})I_i(a_{2k'})}{p_j q_{k'}} \right) (Y_i-\hat{\mu}_{jk'}) \right\}.
\end{aligned} 
\end{equation}

\subsection{Platform SMART allowing adding an additional first-stage treatment}

\begin{figure}
\centering
\includegraphics[width=0.8\textwidth]{"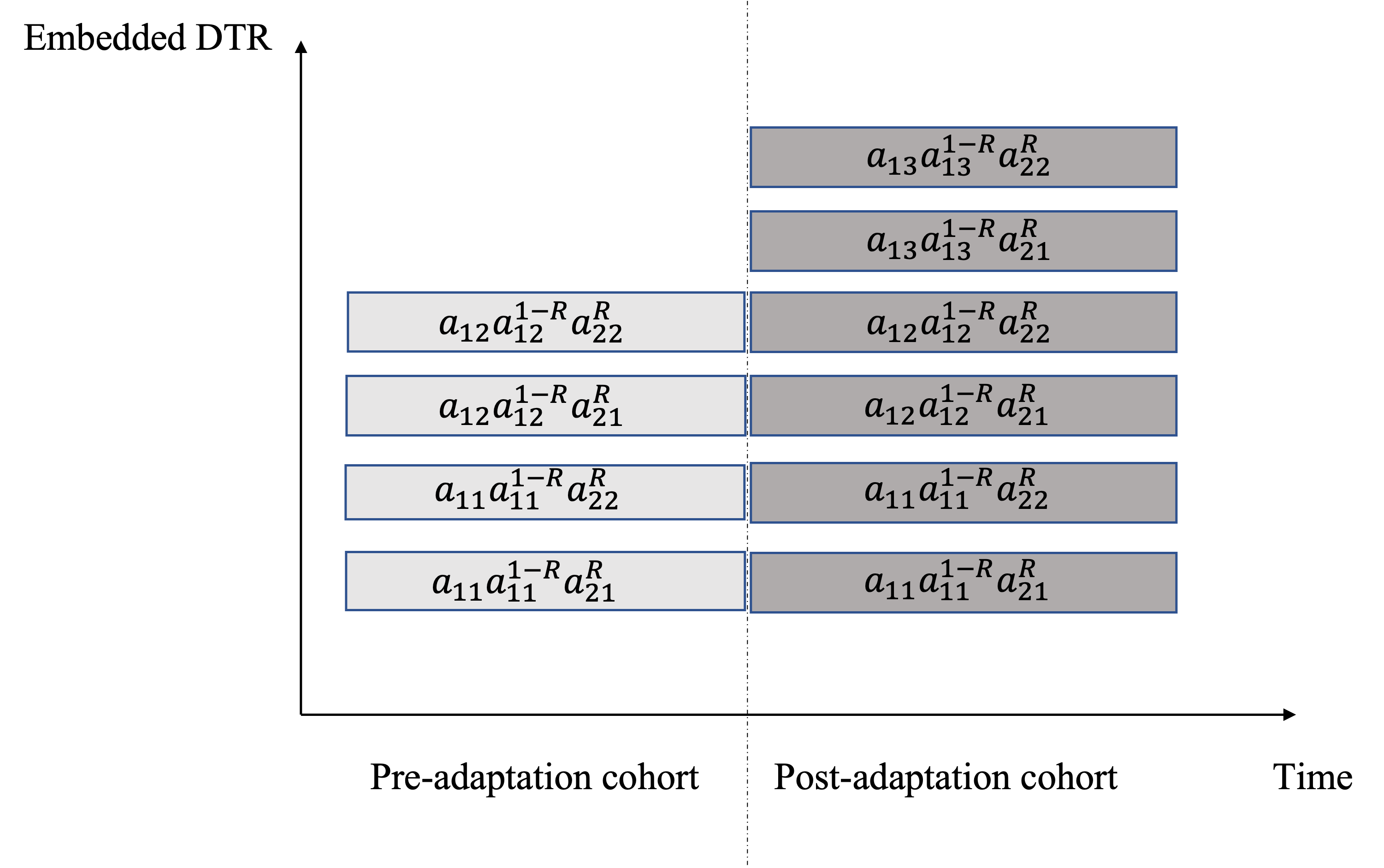"}
\caption{A two-stage platform SMART adding a first-stage treatment $a_{13}$.} \label{figure2}
\end{figure}

Assume that during the trial, an additional first-stage treatment $A_1 = a_{13}$ is available to the ongoing SMART, resulting in a trial with six embedded DTRs (Figure~\ref{figure2}). Using the nomenclature of \citet{marschner2023transparent}, we now have two \textit{concurrently randomized cohorts}, that is, cohorts with the same set of treatments. Each cohort constitutes a mutually exclusive fixed design within the overarching adaptive trial. Henceforth, we shall refer to the concurrently randomized cohort prior to adding treatment $a_{13}$ as the \textit{pre-adaptation cohort} denoted by $C = c_1$ and the concurrently randomized cohort after adding $a_{13}$ as the \textit{post-adaptation cohort} denoted by $C = c_2$, where $C$ is the cohort indicator. Let $n_{1}$ and $n_{2}$ denote the number of participants in $c_1$ and $c_2$, respectively. We denote the randomization probabilities for $a_{1j}$ and $a_{2k}$ in cohort $c_l$, where $l=1,2$, as $p_{jl}$ and $q_{kl}$, respectively. 

The estimated cohort-specific DTR mean is $\hat{\mu}_{jk,l}$, which can be derived based on Eq.~(\ref{mu_1}) using data from cohort $c_l$. The corresponding cohort-specific variance $\widehat{\Var}(\hat{\mu}_{jk,l})$ and covariance $\widehat{\cov}(\hat{\mu}_{jk,l},\hat{\mu}_{jk',l} )$ can be derived based on Eq.~(\ref{eq_var1}--\ref{eq_var4}) with the parameters $n_l$, $p_{jl}$, $q_{kl}$, $\pi_{jl}$, $\mu_{\cdot,l}$, and $\sigma_{\cdot,l}$, representing the cohort-specific sample size, the randomization probabilities for treatment $a_{1j}$ and $a_{2k}$, the response rate for the initial treatment $a_{1j}$, and the expected outcome and variance for the treatment sequences within cohort $C_l$, respectively. We define the set of original embedded DTRs in cohort $c_1$ as $\mathcal{D}_{c_1}$ and the set of newly added embedded DTRs in cohort $c_2$ as $\mathcal{D}_{c_2}$. For DTRs $d_{jk} \in \mathcal{D}_{c_2}$, the non-concurrent and concurrent data refer to individuals assigned to $\mathcal{D}_{c_1}$ in cohort $c_1$ and cohort $c_2$, respectively. Two commonly used methods in the literature on non-concurrent cohorts in platform trials \citep{saville2022bayesian} and historical controls \citep{krotka2023ncc} include the \textit{separate} and \textit{pooling} approaches, which can be extended to platform SMARTs for comparing embedded DTRs. 

\sloppy Adopting the name proposed by \citet{viele2014use}, we refer to the approach that evaluates embedded DTRs using only cohort $c_2$ as the \textit{separate} approach. With estimated DTR means $\hat{\mu}_{jk}^{sep} = \hat{\mu}_{jk,2}$ based on concurrent DTRs in cohort $c_2$, the \textit{separate} approach ensures unbiased estimation of DTR means. Nevertheless, in contrast to traditional settings involving historical controls or external data, the two cohorts within a platform SMART are enrolled under the same protocol, differing primarily in timing and available treatments. Consequently, solely relying on cohort $c_2$ wastes valuable information from cohort $c_1$.

\sloppy The \textit{pooling} approach can be understood as fully utilizing the historical data by estimating DTR means based on data across all cohorts (therefore including data from non-concurrently randomized cohorts). The estimated DTR means for the original embedded DTRs $d_{jk} \in \mathcal{D}_{c_1}$ using the \textit{pooling} approach are denoted as $$\hat{\mu}_{jk}^{pool}= \frac{n_1 \hat{\mu}_{jk,1}+n_2 \hat{\mu}_{jk,2}}{n_1+n_2},$$ with the corresponding variance estimate $$ \widehat{\Var}(\hat{\mu}_{jk}^{pool}) = \frac{n_1^2}{(n_1+n_2)^2} \widehat{\Var}(\hat{\mu}_{jk,1}) + \frac{n_2^2}{(n_1+n_2)^2} \widehat{\Var}(\hat{\mu}_{jk,2}),$$ and the covariance estimate
$$\widehat{\cov}(\hat{\mu}_{jk}^{pool},\hat{\mu}_{jk'}^{pool}) = \frac{n_1^2}{(n_1+n_2)^2} \widehat{\cov}(\hat{\mu}_{jk,1},\hat{\mu}_{jk',1} ) + \frac{n_2^2}{(n_1+n_2)^2} \widehat{\cov}(\hat{\mu}_{jk,2}, \hat{\mu}_{jk',2}).$$ Note that $\hat{\mu}_{jk}^{pool}$ can be seen as a weighted estimator combining estimates from cohort $c_1$ and cohort $c_2$ using weight $n_l/(n_1+n_2)$ for cohort $c_l$. Alternative weightings can also be applied. For example, \citet{roig2022model} demonstrated that estimates from regression models adjusting for time can be expressed as a weighted sum of piecewise treatment effects. 

To ensure that the \textit{pooling} approach produces unbiased estimates of $\mu_{jk}$, the following exchangeability assumption should be satisfied:
\begin{assumption}
\label{assumption1}
    Exchangeability: $Y(d_{jk}) \perp C$, and thus $E[Y(d_{jk}) | c_2] = E[Y(d_{jk}) |  c_1]$.
\end{assumption}

Assumption~(\ref{assumption1}) ensures that the potential outcome under DTR $d_{jk}$ in cohort $c_1$ is the same as that in cohort $c_2$. However, as noted by \citet{lee2020including}, this assumption is unlikely to be fully met in practice due to time effects arising from population shifts, improvements in standard care, and other temporal changes. When such time effects are present, the \textit{pooling} approach may yield biased estimates and lead to invalid inferences.

\section{Bayesian integration G-formula (BIG) estimators.}
\label{section3}

\citet{keil2018bayesian} proposed Bayesian G-formula estimators that can be adopted when the data is sparse due to small sample sizes or when meaningful prior information is available, which in this paper, is the information from the cohort $c_1$. To alleviate the issues of waste of information and potential bias from the \textit{separate} and \textit{pooling} approaches, building upon the work of \citet{keil2018bayesian}, we propose Bayesian integration G-formula (BIG) estimators for platform SMARTs, which allows estimating the DTR means utilizing the non-concurrently randomized cohorts. Note that most of the methods in the literature of historical controls and non-concurrent controls focus on utilizing the prior information for mean or percentage of a single arm. In this paper, we utilize prior information of coefficients in generalized linear models used in BIG estimators. 

Hereafter for simplification, we assume a continuous outcome $Y$ and $X_{1}= \emptyset$ and $X_{2} = R$. Although we present the methodology using continuous outcomes, the proposed approach is applicable to other outcome types. This generality is demonstrated in the application section using binary outcomes. We define the outcome model as $Y \sim f_Y(A_1, R, A_2;\bm{\psi}) + \epsilon$, where $\epsilon \sim N(0, \sigma^2)$ and the response model as $\logit(\pi) = f_R(A_1;\bm{\beta})$, where $\pi$ is the response rate, $\bm{\psi}, \bm{\beta}, \sigma^2$ are model parameters to be estimated. %{\color{blue} Note that, under this model the DTR mean \begin{align*}
%   \mu_{jk}&=E[Y\mid d_{jk},\beta,\psi] \\
%            &=E[Y\mid A_1=a_{1j}, R=1,A_2=a_{2k}]P(R=1\mid A_1=a_{1j})+E[Y\mid A_1=a_{1j}, R=0,A_2=a_{2k}]P(R=0\mid A_1=a_{1j}) \\
%           &=f_Y(A_1=a_{1j}, R=1,A_2=a_{2k};\psi)\pi(A_1=a_{1j};\bm{\beta})+f_Y(A_1=a_{1j}, R=0,A_2=a_{2k};\psi)(1-\pi(A_1=a_{1j};\bm{\beta})),
%\end{align*}
%provided the consistency, exchangeability and positivity assumptions holds here as well. Note that, $\pi(A_1;\bm{\beta})=\frac{e^{f_R(A_1;\beta)}}{1+e^{f_R(A_1;\beta)}}$}
The posterior distribution of the model parameters $\theta = (\bm{\beta}, \bm{\psi}, \sigma)$ based on cohort $c_2$ is expressed as
$$p_{\theta,n_2} (\theta|O_{n_2}) = \mathcal{L}(\theta| O_{n_2} ) p^{(0)}(\theta) = \prod_{i=1}^{n_2} p_Y(Y_i|A_{1i},R_i,A_{2i};\theta) p^{(0)}(\theta),$$
where $\mathcal{L}(\theta| O_{n_2})$ is the likelihood function based on the observed data in cohort $c_2$ and $p^{(0)}(\theta)$ is the prior distribution of $\theta$, which can be specified based on data from cohort $c_1$ \citep{keil2018bayesian}. The predictive distribution of the outcome under DTR $d_{jk}$ is thus denoted as 
\begin{equation}
\label{post1}
%\resizebox{0.9\linewidth}{!}{
    \begin{aligned}
        p_{Y,n_2}(Y|d_{jk}, O_{n_2}) =   &\int_{\theta} 
        \{ p_Y(Y|A_1=a_{1j}, R = 1, A_2=a_{2k}; \bm{\psi}, \sigma) \pi(A_1=a_{1j};\bm{\beta}) \\&\quad+  p_Y(Y|A_1=a_{1j}, R = 0,A_2=a_{2j}; \bm{\psi}, \sigma) (1 - \pi(A_1=a_{1j};\bm{\beta})) \} p_{\theta,n_2}(\theta| O_{n_2}) d\theta,  
    \end{aligned}
%    }
\end{equation}
where $\pi(A_1;\bm{\beta})$ is the response rate under treatment $A_1$ with model parameters $\bm{\beta}$. The mean of DTR $d_{jk}$, $\mu_{jk}$ is thus estimated as $\hat{\mu}_{jk} = E[Y|d_{jk}, O_{n_2}]$, where the expectation is derived from the predictive distribution $p_{Y,n_2}(Y|d_{jk}, O_{n_2})$. In the following subsections we present the detailed steps for the BIG estimator. For simplicity, we illustrate the procedure for obtaining posterior draws of DTR means, as defined in Eq.~(\ref{post1}) using the SMART design shown in \autoref{figure:conventional_SMART}. However, the proposed method is readily extendable to other types of SMART designs and outcomes.

\subsection{Model specification}

The response model for $R$ is specified as $\logit(\pi) = \beta_{0} + \beta_{1} I(a_{12}) + \beta_2 I(a_{13})$. Note that in the SMART shown in \autoref{figure2}, there are only two treatments for cohort $c_1$, so the prior information is provided only for the coefficients $\beta_{0}$ and $\beta_{1}$. This also applies to the subsequent outcome models, where no prior information is provided by cohort $c_1$ for the coefficients of the newly added treatment. The outcome model is specified as
\begin{equation}
%\resizebox{0.9\linewidth}{!}{
    \begin{aligned}
       f_{Y}(A_1,R,A_2;\bm{\psi}) &= R\left\{ \psi_{r,0} +  \psi_{r,1} I(a_{12})+ \psi_{r,2} I(a_{13}) + \psi_{r,3} I(a_{22}) + \psi_{r,4} I(a_{12}) I(a_{22}) + \psi_{r,5} I(a_{13}) I(a_{22}) \right\} \\
       &+ (1-R)\left\{\psi_{nr,0} + \psi_{nr,1} I(a_{12}) + \psi_{nr,2} I(a_{13})\right\}. 
    \end{aligned}
    %}
\end{equation}
We specify both the response model and the outcome model as generalized linear models; however, alternative models can also be utilized in this context. Based on the specified model, the vector of the coefficients for which cohort $c_1$ can provide prior information is $\theta_{s} = (\beta_{0}, \beta_{1}, \psi_{r,0}, \psi_{r,1}, \psi_{r,3}, \psi_{r,4}, \psi_{nr,0}, \psi_{nr,1} )$ with length $B=8$, while the set of coefficients for which cohort $c_1$ cannot provide prior information is $\theta_{ns} = (\beta_{2},  \psi_{r,2}, \psi_{r,5}, \psi_{nr,3})$ with length $G=4$.

\subsection{The prior distributions}

Weakly informative priors can be specified for $\theta_{ns,g}$, which is the $g$-th coefficient in set $\theta_{ns}$, given that there is no prior information. For example, the prior for $\theta_{ns,g}$ can be specified as $p^{(0)}(\theta_{ns,g}) = N(0, 100)$. Alternatively, other weakly informative priors may also be appropriate depending on the specific clinical context. Conversely, for the $b$-th coefficient in set $\theta_{s}$, denoted as $\theta_{s,b}$, prior information from cohort $c_1$ can be leveraged to enhance statistical efficiency. Various prior distributions have been proposed for incorporating non-concurrent data, including log-distance priors, power priors, commensurate priors, and meta-analytic predictive priors \citep{wisniowski2020integrating, salvatore2023bayesian, krotka2023ncc, schmidli2014robust, hobbs2011hierarchical}. However, as pointed out by \citet{hobbs2012commensurate}, in the case of one historical study, the meta-analytic-predictive prior approach is overly sensitive to the priors of the variance parameters. In addition, the power prior approach is computationally intensive, limiting its application in most clinical settings. Given these limitations, we primarily focus on log-distance priors \citep{wisniowski2020integrating, salvatore2023bayesian} and commensurate priors \citep{hobbs2011hierarchical} for the platform SMART.

\subsubsection{The log distance prior}

The log distance prior for $\theta_{s,b}$, which is the $b$-th element in $\theta_s$, is specified as $p^{(0)}_{logdis} (\theta_{s,b}| O_{n_1}) = N \left(\hat{\theta}_{s,b,1}^s, \sqrt{ \max ((\hat{\theta}_{s,b,2} - \hat{\theta}_{s,b,1} )^2, \hat{\sigma}^2_{s,b,1} )} \right) $, where $O_{n_1}$ is the observed data from cohort $c_1$, $\hat{\theta}_{s,b,l}$ and $\hat{\sigma}^2_{s,b,l}$ are the maximum likelihood estimate (MLE) and estimate of its variance based on cohort $c_l$ data \citep{salvatore2023bayesian}. The variance term is defined as the square root of the maximum of the squared difference between the cohort-specific maximum likelihood estimates (MLEs) and the variance of the MLEs from cohort $c_1$, which ensures that the prior distribution of $\theta_{s,b}$ is adaptive, leveraging information from cohort $c_1$ while accounting for differences between the estimates from cohort $c_1$ and cohort $c_2$. Specifically, smaller differences between the cohort $c_1$ and cohort $c_2$ estimates and smaller variance in $\hat{\theta}_{s,b,1}$ result in a tighter and more informative prior distribution centered around $\hat{\theta}_{s,b,1}$. It is to be noted that we have omitted the log-term ($\log(n_1)$ in this case) otherwise present in the denominator of the squared scaling parameter \citep{salvatore2023bayesian,wisniowski2020integrating} as it will reduce the prior variance further. This approach effectively incorporates the uncertainty in the estimates from cohort $c_1$ as well as the heterogeneity between the estimates from the two cohorts, adjusting the prior distribution to account for both factors. Furthermore, the prior for $\sigma$ is denoted as $p^{(0)}(\sigma)$. We assume that the prior distribution for $\sigma$, $\theta_{ns}$, and $\theta_s$ are independent such that the joint prior distribution for $\sigma$ and $\theta_s$, and $\theta_{ns}$ is expressed as:
$$p^{(0)}_{logdis}(\theta_{s}, \theta_{ns}, \sigma| O_{n_1}) = \prod_{b=1}^B p^{(0)}_{logdis} (\theta_{s,b}| O_{n_1}) \prod_{g=1}^G p^{(0)}(\theta_{ns,g}) p^{(0)}(\sigma).$$
\subsubsection{The commensurate priors}

Another prior distribution is the commensurate prior, which incorporates a parameter to quantify the commensurability between cohort-specific estimators. The conditional prior distribution of $\theta_{s,b}$ given the cohort $c_1$ estimate $\theta_{s,b,1}$ and the commensurability parameter $\tau$ is defined as 
$$p_{comP}(\theta_{s,b} | \theta_{s,b,1}, \tau) = N(\theta_{s,b,1}, 1/\tau).$$ The joint prior for $\theta_{s}$, $\theta_{ns}$ and $\sigma$ is expressed as
\begin{equation}
\label{CP_prior}
    \begin{aligned}
       p^{(0)}_{comP}(\theta_{s}, \theta_{ns}, \sigma| O_{n_1}) = \int_{\tau} \int_{\theta_{s,1}} p^{(0)}(\theta_{s,1}) p^{(0)}(\tau) p^{(0)}(\theta_{ns}) p^{(0)}(\sigma) \mathcal{L}(\theta_{s,1}, \sigma|O_{n_1})  p_{comP}(\theta_{s}|\theta_{s,1}, \tau) d\theta_{s,1} d \tau,   
    \end{aligned}
\end{equation}
where $\theta_{s,1}$ is the estimate of set $\theta_s$ based on cohort $c_1$, and $p^{(0)}(\theta_{s,1})$, $p^{(0)}(\tau)$, $p^{(0)}(\theta_{ns}), p^{(0)}(\sigma)$ are the prior distributions for $\theta_{s,1}$, $\tau$, $\theta_{ns}$, and $\sigma$. The value of $\tau$ controls the prior precision of $\theta_{s}$ given the estimates based on $\theta_{s,1}$. Higher values of $\tau$ indicate greater borrowing of information from the non-concurrent data, whereas $\tau$ approaching zero represents no borrowing of non-concurrent data. As the parameters $\theta_{s,1}$ and $\tau$ are not of interest for comparing DTRs in cohort $c_2$, we integrate them over to get the posterior distributions of DTR means.

An alternative commensurate prior, known as the mixed commensurate prior proposed by \citet{hobbs2011hierarchical}, has been shown to perform comparably to standard commensurate priors while avoiding the need to compute posterior draws of the commensurate parameter. We begin by specifying a set of fixed values for $\bm{\tau} = (\tau_1,\dots, \tau_H), h = 1, \dots, H$ along with their corresponding weights $\bm{w} = (w_1, \dots, w_H)$. The joint prior distribution for $\theta_{s}$, $\theta_{ns}$, and $\sigma$ is
\begin{equation}
\label{MCP_prior}
    \begin{aligned}
       p^{(0)}_{commP}(\theta_{s}, \theta_{ns}, \sigma | O_{n_1}, \bm{\tau}, \bm{w}) = 
       \sum_{h=1}^H w_h \int_{\theta_{s,1}}  p^{(0)}(\theta_{s,1}) p^{(0)}(\theta_{ns}) p^{(0)}(\sigma) \mathcal{L}(\theta_{s,1}, \sigma|O_{n_1})  p_{comP}(\theta_{s}|\theta_{s,1}, \tau_h) d\theta_{s,1}.
    \end{aligned}
\end{equation}

Each weight controls the relative importance of its corresponding commensurate parameters. For example, in the subsequent simulation section, we adopt two sets of commensurate parameters ($H=2$): 
 $\tau_{1} = 0.1$ and $\tau_{2} = 20$. In addition, equal weights are given for each set, i.e., $w_1 = w_2 = 0.5$. The first set of scaling parameters gives relatively lower precision to the prior distribution based on cohort $c_1$ estimates, while the second set of scaling parameters corresponds to the prior information of high commensurabilities between two cohorts. Including the second set of scaling parameters in addition to the first set facilitates a `partial borrowing' by utilizing relatively larger scaling parameters.

\subsection{Posterior distributions of DTR means}

The posterior distributions of the model parameters are denoted as $p_{\theta_s, \theta_{ns}, \sigma, n_2}(\theta_s, \theta_{ns}, \sigma|O_{n_2}) = \mathcal{L}(\theta_s, \theta_{ns}, \sigma|O_{n_2}) p^{(0)}(\theta_s,\theta_{ns},\sigma|O_{n_1})$, where $p^{(0)}(\theta_s,\theta_{ns},\sigma|O_{n_1})$ can be specified based on the log distance prior, commensurate prior, or mixed commensurate prior. As mentioned earlier, we estimate the DTR means using the Bayesian G-formula algorithm \citep{keil2018bayesian} which consists of generating observations from the posterior predictive distribution under a particular DTR. Towards this, we generate $M$ samples from the posterior distribution $p_{\theta_s, \theta_{ns}, \sigma, n_2}(\theta_s, \theta_{ns}, \sigma|O_{n_2})$. For every $m$-th draw $(\theta_s^{(m)},\theta_{ns}^{(m)}, \sigma^{(m)})$, $m=1,2,\dots,M$, a population of size $N$ is generated, and the initial treatment is assigned according to the DTR of interest $d_{jk}$. The outcomes $Y_i^{(m)}$ and responses to initial treatment $R_i^{(m)}$ are generated according to the model
\begin{align*}
    R_i^{(m)} &\sim Bernoulli(\pi(A_1=a_{1j};\beta^{(m)}), \\
    \text{and, }Y_i^{(m)}\mid R_i^{(m)} &\sim N(f_Y(A_1=a_{1j},R_i,A_2=a_{2k};\psi^{(m)}),\sigma^{2(m)}),
\end{align*}
$i=1,2,\dots,N$. Thus, we have the mean response corresponding to the $m$th draw, $\mu_{jk}^{(m)}=\frac{1}{N}\sum_{i=1}^NY_i^{(m)}$, and the collection $\{\mu_{jk}^{(m)}\}_{m=1}^M$ approximates the posterior distribution of the DTR mean $\mu_{jk}$.

\subsection{The estimands of interest}

The sample mean of the posterior draws is taken to be the estimated DTR mean $\hat{\mu}_{jk}$, and the corresponding variance is the standard estimate of $\hat{\mu}_{jk}$, i.e., 
$$\hat{\mu}_{jk}=\frac{1}{M}\sum_{m=1}^M \mu_{jk}^{(m)} ~ \text{and} ~ \widehat{Var}(\hat{\mu}_{jk}) = \frac{\sum_{m=1}^M (\mu^{(m)}_{jk}-\hat{\mu}_{jk})^2}{M-1}.$$

The variance of the estimated difference between two DTR means $\hat{\mu}_{jk} - \hat{\mu}_{j'k'}$ is $\widehat{Var}(\hat{\mu}_{jk} - \hat{\mu}_{j'k'}) = \frac{\sum_{m=1}^M \{\mu^{(m)}_{jk} - \mu^{(m)}_{j'k'}-(\hat{\mu}_{jk} - \hat{\mu}_{j'k'})\}^2}{M-1}$.

\section{Simulation study}
\label{section4}

In our simulation study, we focus on the SMART in Figure~\ref{figure2}. To explore how the timing of introducing a new treatment influences the design's operating characteristics, we set the ratio of $n_1$ to $n$, denoted as $r$, to different values, i.e., $r = 0.3, 0.5, 0.7$ for each design, where $n$ is the pre-specified sample size for the original SMART and $n_1$ denotes the number of cohort $c_1$, i.e., the number of patients recruited at the time when treatment $a_{13}$ is added. We consider different sample sizes $n = 500, 1000, 1500$. 

Globally, we assume equal randomization probabilities in cohort $c_1$, that is, $p_{j1}=0.5$ and $q_{k1}=0.5$. However, for cohort $c_2$, the allocation of sample size for the initial treatment is skewed towards treatment $a_{13}$ to ensure balanced sample sizes for each treatment. Specifically, the total sample size for $a_{11}$ and $a_{12}$ is $n/2$. The sample size for $a_{11}$ and $a_{12}$ in cohort $c_1$ is $n_1/2$. For cohort $c_2$, the sample size for $a_{11}$, $a_{12}$, and $a_{13}$ should be $n/2-n_1/2$, $n/2-n_1/2$ and $n/2$, respectively, to ensure that the total sample size for each treatment is $n/2$. Additionally, we assume equal randomization probabilities for the second-stage treatments for cohort $c_2$, i.e., $q_{k2}=0.5$. We generate the response rate $R$ and the continuous outcome $Y$ under five scenarios detailed in Table~\ref{table_1}. 

\begin{table}[htb]
    \centering
    \caption{Parameters for generating the response rate $R$ and the outcome $Y$ under five scenarios. }
    \label{table_1}
    \resizebox{\textwidth}{!}{\begin{tabular}{lccccccccccc}
\toprule
&   & \multicolumn{2}{c}{Scenario 1} & \multicolumn{2}{c}{Scenario 2} & \multicolumn{2}{c}{Scenario 3} & \multicolumn{2}{c}{Scenario 4} & \multicolumn{2}{c}{Scenario 5} \\
& & $c_1$ & $c_2$ &$c_1$ & $c_2$ & $c_1$ & $c_2$ & $c_1$ & $c_2$ & $c_1$ & $c_2$ \\
\midrule  
\multirow{3}{4em}{Response rate} 
& $a_{11}$ & 0.5 & 0.5 & 0.4 & 0.4 & 0.4 & 0.4 & 0.4 & 0.5 & 0.4 & 0.5 \\
& $a_{12}$ & 0.5 & 0.5 & 0.5 & 0.5 & 0.5 & 0.5 & 0.5 & 0.6 & 0.5 & 0.6 \\
& $a_{13}$ & 0.5 & 0.5 & 0.6 & 0.6 & 0.6 & 0.6 & 0.6 & 0.7 & 0.6 & 0.7 \\
\multirow{9}{4em}{Treatment sequence means}

& $\mu_{a_{11} a_{11}}$ & 15.0 & 15.0 & 15.0 & 15.0 & 15.0 & 15.0 & 16.0 & 18.0 & 15.0 & 17.0\\
 & $\mu_{a_{11} a_{21}}$ & 17.0 & 17.0 & 17.0 & 17.0 & 17.0 & 17.0 & 19.0 & 21.0 & 17.0 & 19.0\\
 & $\mu_{a_{11} a_{22}}$ & 17.0 & 17.0 & 18.0 & 18.0 & 18.0 & 18.0 & 18.0 & 20.0 & 18.0 & 20.0\\
 & $\mu_{a_{12} a_{12}}$ & 15.0 & 15.0 & 19.0 & 19.0 & 19.0 & 19.0 & 19.0 & 21.0 & 19.0 & 21.0\\
 & $\mu_{a_{12} a_{21}}$ & 17.0 & 17.0 & 18.0 & 18.0 & 18.0 & 18.0 & 18.0 & 20.0 & 18.0 & 20.0\\
 & $\mu_{a_{12} a_{22}}$ & 17.0 & 17.0 & 16.0 & 16.0 & 16.0 & 16.0 & 16.0 & 18.0 & 16.0 & 18.0\\
 & $\mu_{a_{13}a_{13}}$ & 15.0 & 15.0 & 16.0 & 16.0 & 20.0 & 20.0 & 20.0 & 22.0 & 21.0 & 22.0\\
 & $\mu_{a_{13}a_{21}}$ & 17.0 & 17.0 & 16.0 & 16.0 & 19.0 & 19.0 & 19.0 & 21.0 & 19.0 & 20.0\\
 & $\mu_{a_{13}a_{22}}$ & 17.0 & 17.0 & 17.0 & 17.0 & 17.0 & 17.0 & 17.0 & 19.0 & 17.0 & 18.0\\
 \hline
\multirow{6}{4em}{DTR means} 
& $\mu_{11}$ & 16.0 & 16.0 & 15.8 & 15.8 & 15.8 & 15.8 & 17.2 & 19.5 & 15.8 & 18.0\\
 & $\mu_{12}$ & 16.0 & 16.0 & 16.2 & 16.2 & 16.2 & 16.2 & 16.8 & 19.0 & 16.2 & 18.5\\
 & $\mu_{21}$ & 16.0 & 16.0 & 18.5 & 18.5 & 18.5 & 18.5 & 18.5 & 20.4 & 18.5 & 20.4\\
 & $\mu_{22}$ & 16.0 & 16.0 & 17.5 & 17.5 & 17.5 & 17.5 & 17.5 & 19.2 & 17.5 & 19.2\\
 & $\mu_{31}$ & 16.0 & 16.0 & 16.0 & 16.0 & 19.4 & 19.4 & 19.4 & 21.3 & 19.8 & 20.6\\
 & $\mu_{32}$ & 16.0 & 16.0 & 16.6 & 16.6 & 18.2 & 18.2 & 18.2 & 19.9 & 18.6 & 19.2\\
 \hline
Assumption 1 & & \multicolumn{2}{c}{satisfied} & \multicolumn{2}{c}{satisfied} & \multicolumn{2}{c}{satisfied} & \multicolumn{2}{c}{not satisfied} & \multicolumn{2}{c}{not satisfied} \\
\hline
    \end{tabular}}
\end{table}

Specifically, in scenario 1, the null hypothesis is true; while scenarios 2--5 are aimed at evaluating the empirical power of the candidate approaches with different optimal DTRs and time effect patterns. In scenarios 2 and 3, there is no time effect, satisfying Assumption 1. The optimal DTR is $d_{21}$ in scenario 2 and $d_{31}$ in scenario 3. Scenarios 4 and 5 are aimed at evaluating the impact of different patterns of time effects. Fixed time effects are applied to all treatment sequence means and response rates in scenario 4. Under scenario 5, different time effects are applied to treatment sequence means. 

We use the three different approaches described earlier to estimate DTR means in cohort $c_2$: 1) the \textit{separate} approach using only cohort $c_2$ data; 2) the pooling approach combining estimates from the two cohorts; and 3) the BIG approaches using different prior distributions for the parameters shared by the two cohorts. Specifically, for the BIG approaches, we consider the weakly informative priors denoted as BIGweak, the log distance prior denoted as BIGlogdis, and the commensurate priors and mixed commensurate priors denoted as BIGcomP and BIGcommP, with two sets of scaling parameters $\tau_{1} = 0.1$, and $\tau_{2} = 20$ and equal weights for each set, i.e., $w_1 = w_2 = 0.5$. 

For each scenario, we simulate $1,000$ trials to evaluate the performance of each approach in terms of the probability of identifying the true optimal DTR $P(\hat{d}^{\ast}=d^{\ast})$, bias $E[(\mu_{11}-\mu_{31}) - (\hat{\mu}_{11}-\hat{\mu}_{31})]$, empirical variance $\Var((\hat{\mu}_{11}-\hat{\mu}_{31}))$, coverage rate $P(\mu_{11}-\mu_{31} \in \hat{CI} )$, and mean squared error (MSE) $E[((\mu_{11}-\mu_{31}) - (\hat{\mu}_{11}-\hat{\mu}_{31}))^2]$ in estimating $\mu_{11}-\mu_{31}$, i.e., the difference between DTR means for $d_{11}$ and $d_{31}$ in cohort $c_2$ across simulation replicates, where $\widehat{CI}$ is the estimated confidence/credible interval for $\mu_{11}-\mu_{31}$.

\subsection{Simulation results}

\begin{figure}[htb]
\centering
\includegraphics[width=0.8\textwidth]{"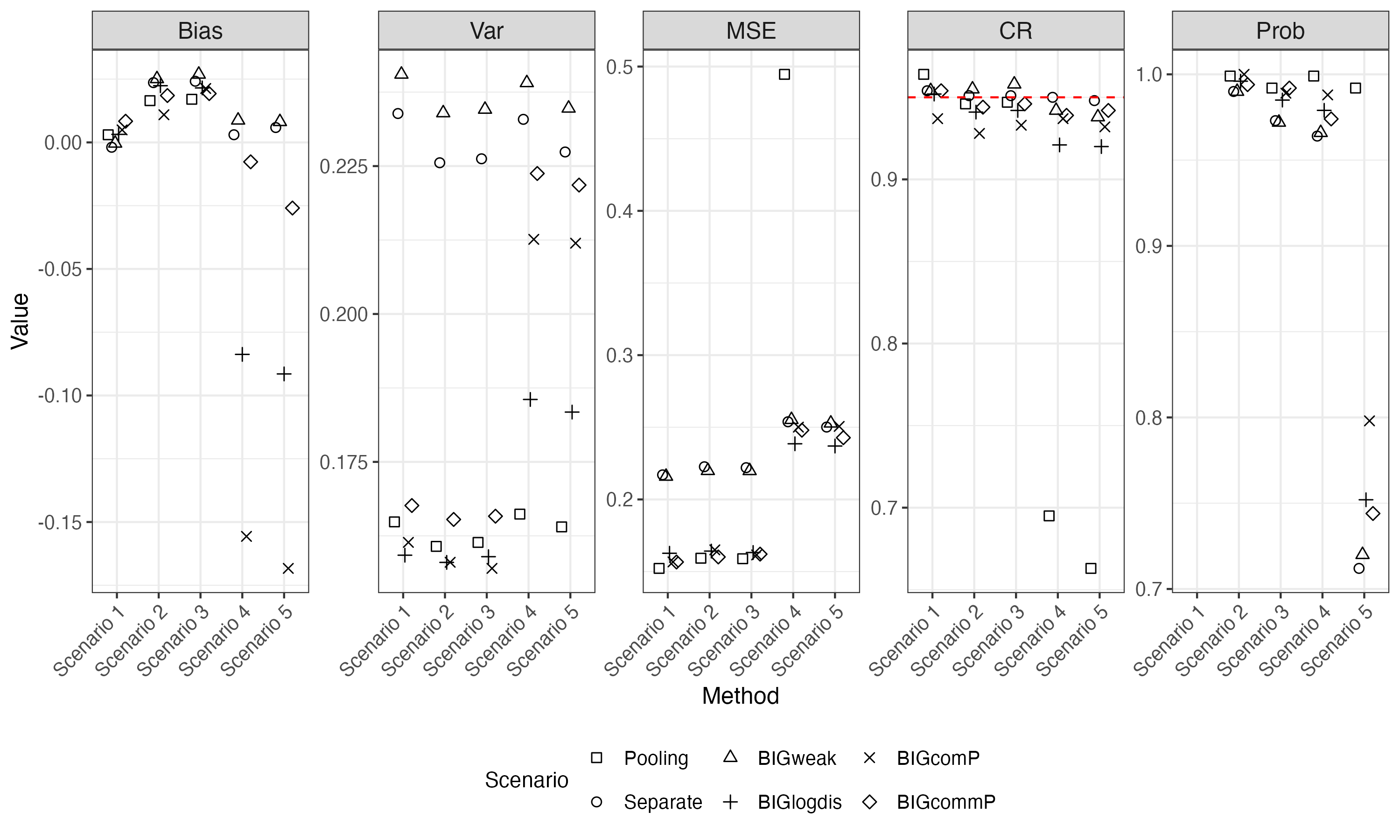"}
\caption{Simulation results with $n=1000$ and $r=0.5$. `Bias', `Var', `MSE', and `CR' represent the corresponding metrics (bias, variance, mean squared error, and cover rate) in terms of estimating $\mu_{11}-\mu_{31}$, i.e., the difference in expected outcome for DTRs $d_{11}$ and $d_{31}$ at cohort $c_2$. `Prob' represents the probability of identifying the true optimal DTR. BIGweak, BIGlogdis, BIGcomP and BIGcommP are the Bayesian integration G-formula (BIG) approaches with weakly informative priors, log distance priors, commensurate priors, and mixed commensurate priors.}
\label{figure3}
\end{figure}

Figure~\ref{figure3} summarizes the simulation results comparing the candidate approaches with sample size $n=1000$ and $r=0.5$. For better illustration, the results with absolute bias higher than 0.4, MSE higher than 0.5, and coverage rate smaller than 0.7 are not presented in the figure. In addition, we ignore scenario 1 for the probability of identifying the true optimal DTR given that there is no optimal DTR under this setting. 

In scenario 1, all approaches yield nearly unbiased estimates. However, in scenarios 2 and 3 without time effects, BIGlogdis, BIGcomP, and BIGcommP demonstrate lower bias compared to the  \textit{separate} and BIGweak approaches. In contrast, scenarios 4 and 5 reveal substantial bias in the \textit{pooling} approach due to the time effect (exceeding an absolute bias of 0.4 thus omitted from the figure), followed by the \textit{separate} and BIGweak approaches. With respect to variance, the \textit{separate} and BIGweak approaches consistently show the highest variability across all scenarios. In Scenarios 1–3, the pooling and BIGcommP approaches yield lower variance than BIGlogdis and BIGcomP, while this pattern reverses in scenarios 4 and 5. The MSE follows a similar trend: the \textit{separate} and BIGweak approaches have higher MSE in scenarios 1-3. whereas the \textit{pooling} approach generates the highest MSE followed by the \textit{separate} and BIGweak approaches in scenarios 4 and 5. Coverage rates are close to nominal levels -- 0.95 -- for all approaches except for scenarios 4 and 5, where the \textit{pooling} approach generates significantly lower coverage rates due to the biased estimates. Regarding the probability of identifying the true optimal DTR, the BIGlogdis, BIGcomP, and BIGcommP approaches outperform the \textit{separate} and BIGweak approaches in all scenarios. Although the \textit{pooling} approach performs well in scenarios 4 and 5, its performance is driven by biased estimates that favor the true optimal DTR.

\iffalse
In summary, using the cohort $c_1$ data as the prior distributions for analyzing cohort $c_2$ data can improve efficiency compared with the \textit{separate} and BIGweak approaches and improve accuracy compared with the \textit{pooling} approach.\fi

\iffalse
\begin{figure}[htb]
\centering
\includegraphics[width=\textwidth]{"./Graphs/0.5.png"}
\caption{Boxplot for the estimated difference $\mu_{11}-\mu_{31}$ with $r=0.5$. }
\label{figure4}
\end{figure}

Figure~\ref{figure4} shows the boxplot of the estimated difference $\mu_{11}-\mu_{31}$ with $r=0.5$ and different sample sizes $500, 1000, 1500$. \fi 

We also provide the simulation results for other scenarios with $r=0.3, 0.5, 0.7$ and sample sizes $n=500, 1000, 1500$ in Figures \ref{fig:sim1} to \ref{fig:sim8} in the appendix. One can observe that as $r$ increases, the probability of identifying the true optimal DTR decreases for all of these approaches in most scenarios because higher values of $r$ correspond to less data in cohort $c_2$. Furthermore, as the value of $r$ increases, the bias for the \textit{pooling} approach in scenarios 4 and 5 also exhibits an increasing trend. These findings indicate that the later the new treatment is added, the more biased estimates the \textit{pooling} approach will produce. Similar patterns are observed for simulations with different sample sizes.

In summary, the simulation results demonstrate that the \textit{pooling} approach can efficiently estimate the difference in DTR means without bias when there are no time effects (scenarios 1--3); however, when there are time effects (scenarios 4 and 5), it may generate biased estimates. The \textit{separate} and BIGweak approaches can generate unbiased estimates in all scenarios but with relatively higher variance and MSE. The BIG approaches BIGlogdis, BIGcomP, and BIGcommP balance the trade-off between enhancing statistical efficiency and maintaining unbiased estimates by taking into account the commensurabilities between the two cohorts, i.e., the difference between the cohort-specific estimates. 

\section{Application}
\label{section5}

We demonstrate the application of the proposed methods within the SNAP trial, a multifactorial Bayesian adaptive platform trial for patients with Staphylococcus aureus bloodstream infection \citep{tong2022staphylococcus}. The SNAP trial aims to simultaneously evaluate multiple treatments across different treatment domains and participant subgroups, including methicillin-susceptible (S. aureus, MSSA), penicillin-susceptible (S. aureus, PSSA), and methicillin-resistant S. aureus (MRSA) \citep{mahar2023blueprint}. Given the variability in the characteristics of the infecting S. aureus bacterium, as well as substantial heterogeneity in diagnostic and therapeutic practices, recruiting sufficient patients into multiple fixed-size trials to draw robust conclusions about multiple treatments presents significant challenges. The SNAP trial addresses these challenges using an adaptive platform design, enabling the  evaluation of multiple treatments within a unified trial framework.

Specifically, in the original SNAP trial, there are three treatment domains: 1) a backbone antibiotic domain comprising antibiotics specific to the infecting \textit{S. aureus} bacterium; 2) an adjunctive antibiotic domain comprising antibiotics applicable to all infecting S. aureus bacteria, and 3) an early oral switch domain comprising standard care and an early switching of antibiotics from intravenous to oral delivery routes for those who demonstrate a desirable response to the initial randomised treatment. New treatments can be added for each specific treatment domain and/or subgroups of patients during the trial and inferior treatments can be dropped based on the interim analytical results. The original SNAP trial protocol specifies multiple different endpoints for analysis, including binary, continuous, time-to-event, and ordinal endpoints. In this paper we only consider this binary outcome, which is the all-cause mortality at 90 days after platform entry. A detailed clinical description and rationale for the SNAP trial can be found at \citet{tong2022staphylococcus}, with technical details and simulation results found in \citet{mahar2023blueprint}. 

While the SNAP trial aims to identify domain-specific treatment efficacies, rather than DTRs, it can also be viewed as a SMART design (i.e. responders to treatment are randomised to receive early oral switch, or not).  For our demonstration, we only consider the adjunctive antibiotic as the first-stage treatment and the early oral switch as the second-stage treatment and only consider adult patients with MSSA infection. Given that the SNAP trial is still ongoing, we demonstrate our proposed approaches based on the parameters used to simulate the multiple virtual trials to evaluate the operating characteristics of the proposed trial design \citep[see][for details of the simulation mechanism]{mahar2023blueprint}. The randomization scheme, however, is similar to our simulation study in Section~\ref{section4}. 

Similar to the simulation study in Section~\ref{section4}, we include three scenarios regarding the generation of binary outcomes: scenario 1) the null scenario where all the embedded DTRs have the same means; scenario 2) the alternative scenario without time effects where the optimal DTR is $d_{31}$; and scenario 3) the alternative scenario with time effects where the optimal DTR is $d_{31}$. The details of the parameters are provided in Table~\ref{table_2}. Specifically, the mortality risks for the reference initial treatment $a_{11}$ among the non-responders and responders are assumed to be $16.8\%$ and $15.0\%$. In scenario 1, we assume that mortality risks are consistent across all DTRs. In scenario 2, the optimal DTR is $d_{31}$, with no time effect. Scenario 3 differs from scenario 2 in that the former has a time effect with an odds ratio of $1.5$ when comparing cohort $c_1$ with cohort $c_2$. 

\begin{table}[htb]
    \centering
    \caption{Parameters for generating the response rate $R$ and the outcome $Y$ under three scenarios for the SNAP trial. }
    \label{table_2}
    \resizebox{0.7\textwidth}{!}{\begin{tabular}{lccccccccccc}
\toprule
&   & \multicolumn{2}{c}{Scenario 1} & \multicolumn{2}{c}{Scenario 2} & \multicolumn{2}{c}{Scenario 3}\\
& & $c_1$ & $c_2$ &$c_1$ & $c_2$ & $c_1$ & $c_2$ \\
\midrule  
\multirow{3}{4em}{Response rate} 
& $a_{11}$ & 0.550 & 0.550 & 0.550 & 0.550 & 0.550 & 0.550\\
 & $a_{12}$ & 0.500 & 0.500 & 0.500 & 0.500 & 0.500 & 0.500\\
 & $a_{13}$ & 0.450 & 0.450 & 0.450 & 0.450 & 0.450 & 0.450\\
\multirow{9}{4em}{Treatment sequence means}

& $\mu_{a_{11}a_{11}}$ & 0.168 & 0.168 & 0.168 & 0.168 & 0.168 & 0.230\\
 & $\mu_{a_{11}a_{21}}$ & 0.150 & 0.150 & 0.150 & 0.150 & 0.150 & 0.210\\
 & $\mu_{a_{11}a_{22}}$ & 0.150 & 0.150 & 0.160 & 0.160 & 0.160 & 0.220\\
 & $\mu_{a_{12}a_{12}}$ & 0.168 & 0.168 & 0.200 & 0.200 & 0.200 & 0.270\\
 & $\mu_{a_{12}a_{21}}$ & 0.150 & 0.150 & 0.160 & 0.160 & 0.160 & 0.220\\
 & $\mu_{a_{12}a_{22}}$ & 0.150 & 0.150 & 0.190 & 0.190 & 0.190 & 0.260\\
 & $\mu_{a_{31}a_{31}}$ & 0.168 & 0.168 & 0.140 & 0.140 & 0.140 & 0.200\\
 & $\mu_{a_{31}a_{21}}$ & 0.150 & 0.150 & 0.120 & 0.120 & 0.120 & 0.170\\
 & $\mu_{a_{31}a_{22}}$ & 0.150 & 0.150 & 0.130 & 0.130 & 0.130 & 0.240\\
 \hline
\multirow{6}{4em}{DTR means} 
& $\mu_{11}$ & 0.158 & 0.158 & 0.158 & 0.158 & 0.158 & 0.219\\
 & $\mu_{12}$ & 0.158 & 0.158 & 0.164 & 0.164 & 0.164 & 0.224\\
 & $\mu_{21}$ & 0.159 & 0.159 & 0.180 & 0.180 & 0.180 & 0.245\\
 & $\mu_{22}$ & 0.159 & 0.159 & 0.195 & 0.195 & 0.195 & 0.265\\
 & $\mu_{31}$ & 0.160 & 0.160 & 0.131 & 0.131 & 0.131 & 0.187\\
 & $\mu_{32}$ & 0.160 & 0.160 & 0.136 & 0.136 & 0.136 & 0.218\\
\hline
    \end{tabular}}
\end{table}

We consider sample sizes $n = 2000, 3000, 4000$ and $r = 0.3, 0.5, 0.7$ and generate $1000$ simulation replicates for each scenario. The results for the scenario with $r=0.5$ and $n=2000$ are shown in Figure~\ref{figure4} and the results for other scenarios are provided in Figures \ref{fig:app1} to \ref{fig:app8} in the Appendix. In scenarios 1 and 2, all approaches exhibit minimal bias and the coverage rates are close to the nominal level. However, the variance and MSE of \textit{pooling}, BIGlogdis, BIGcomP, and BIGcommP approaches are lower than that of the \textit{separate} and BIGweak approaches. In scenario 3, the \textit{separate} and BIGweak approaches generate the lowest bias compared with other approaches, followed by BIGlogdis, BIGcommP, and BIGcomP. The \textit{pooling} approach has the highest bias, the lowest coverage rate, and the lowest probability of identifying the optimal DTR compared with other approaches due to the time effect. As expected, the BIG approaches, BIGlogdis, BIGcomP, and BIGcommP, can improve the efficiencies demonstrated by the relatively lower empirical variance and MSE compared with the \textit{separate} and BIGweak approaches. The results for other scenarios provided in the Appendix show similar patterns. 

\begin{figure}[!ht]
\centering
\includegraphics[width=0.8\textwidth]{"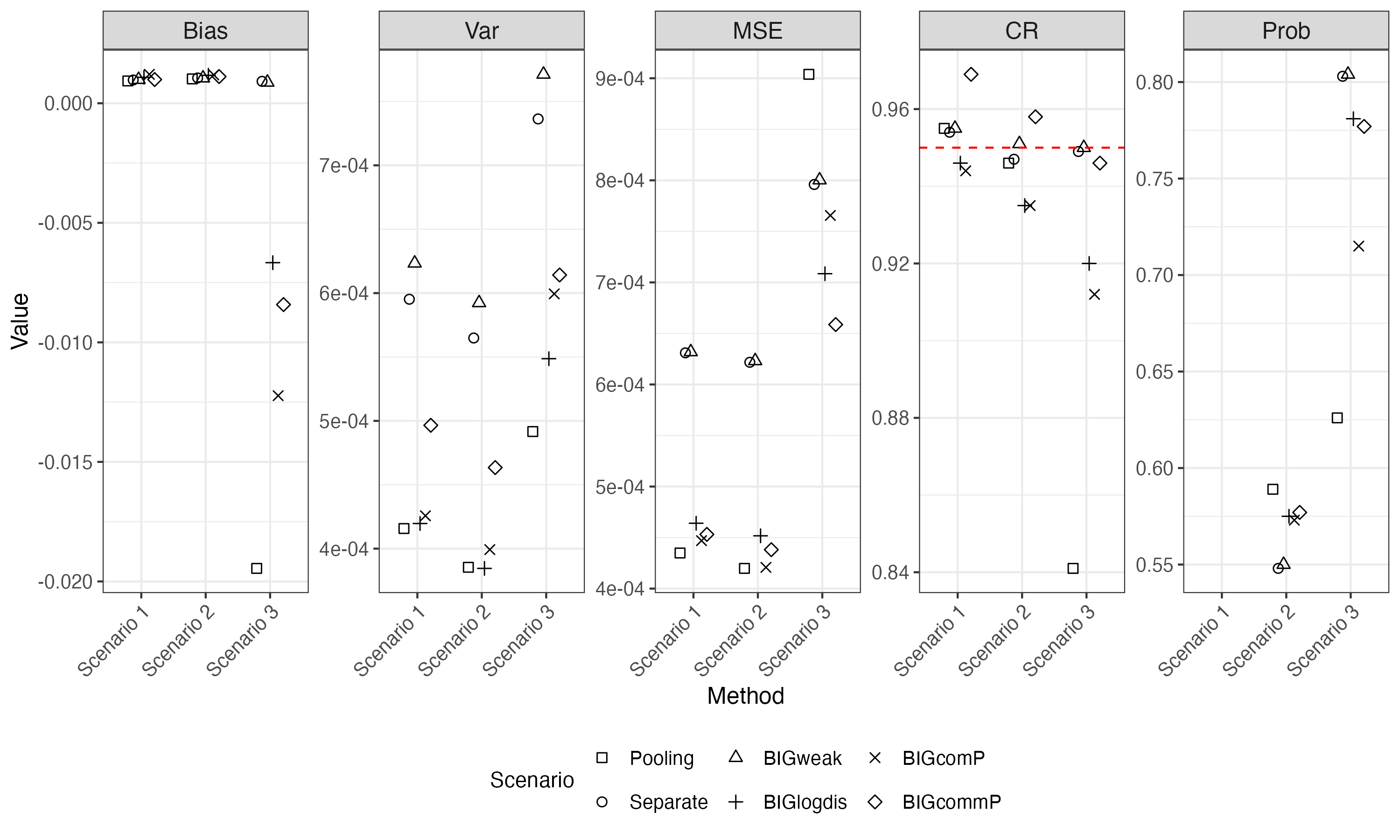"}
\caption{Application results with $n=2000$ and $r=0.5$. `Bias', `Var', `MSE', and `CR' represent the corresponding metrics (bias, variance, mean squared error, and cover rate) in terms of estimating $\mu_{11}-\mu_{31}$, i.e., the difference in expected outcome for DTRs $d_{11}$ and $d_{31}$ at cohort $c_2$. `Prob' represents the probability of identifying the true optimal DTR. BIGweak, BIGlogdis, BIGcomP and BIGcommP are the Bayesian integration G-formula (BIG) approaches with weakly informative priors, log distance priors, commensurate priors, and mixed commensurate priors. }
\label{figure4}
\end{figure}

\section{Discussion}
\label{section6}

Conventional SMART designs often require a longer duration than single-stage RCTs, increasing the likelihood that new treatments may emerge during the trial period. In this paper, we consider a flexible SMART design (which we refer to as platform SMART design) where a new initial treatment is allowed to be added during the trial based on the SNAP trial. By integrating features of platform trials and SMART designs, the platform SMART has the potential to accelerate trial progress and enable earlier availability of effective DTRs for patients. We consider three analytical approaches: the \textit{separate}, \textit{pooling}, and Bayesian integration G-formula (BIG) approaches for platform SMARTs and conduct simulation studies to evaluate the performance of these approaches under various scenarios. 

The \textit{pooling} approach shares similarities with fully utilizing historical control data to increase statistical efficiency. However, it may introduce bias in the presence of time effects, which are unavoidable in SMART designs. On the other hand, the \textit{separate} and BIGweak approaches compare DTRs only using the cohort $c_2$ data, aiming to minimize potential bias introduced by time effects. However, valuable information may be wasted when ignoring the cohort $c_1$ data. The proposed BIG approaches with different priors provide a compromise that alleviates potential biases due to potential time effects while improving efficiency of the estimates. 

We compare these candidate approaches for platform SMARTs through a simulation study. The simulation results indicate that, when there are no time effects, the pooling approach outperforms the other approaches in terms of the variance and the probability of identifying the optimal DTR. However, when there are time effects, BIGlogdis, BIGcomP and BIGcommP generate lower variance compared to the \textit{separate} and the BIGweak approaches and lower bias compared to the pooling approach. 

One limitation of this paper is that we focus only on cases with a single newly added initial treatment. However, multiple treatments may be available at either the initial or the second stage. Additionally, we employ generalized linear regression for the response and the outcome models, while more flexible modeling approaches could be used to mitigate potential model misspecification. Exploring these extensions remains an area for future research.

\section*{Acknowledgments}
Xinru Wang would like to acknowledge her PhD scholarship from the Duke-NUS Medical School, Singapore. Bibhas Chakraborty would like to acknowledge support from the grant MOE-T2EP20122-0013 from the Ministry of Education, Singapore, and the Cross-faculty grant 25-2516-A0001 from the National University of Singapore.

\section*{Conflict of Interest}
The authors declare no conflict of interest. 

%\newpage

%\bibliographystyle{plainnat}
\bibliographystyle{apalike}
\bibliography{refer.bib}

\appendix
\section*{Appendix}
\section{Simulation results}

\begin{figure}[!ht]
\centering
\includegraphics[width=0.8\textwidth]{"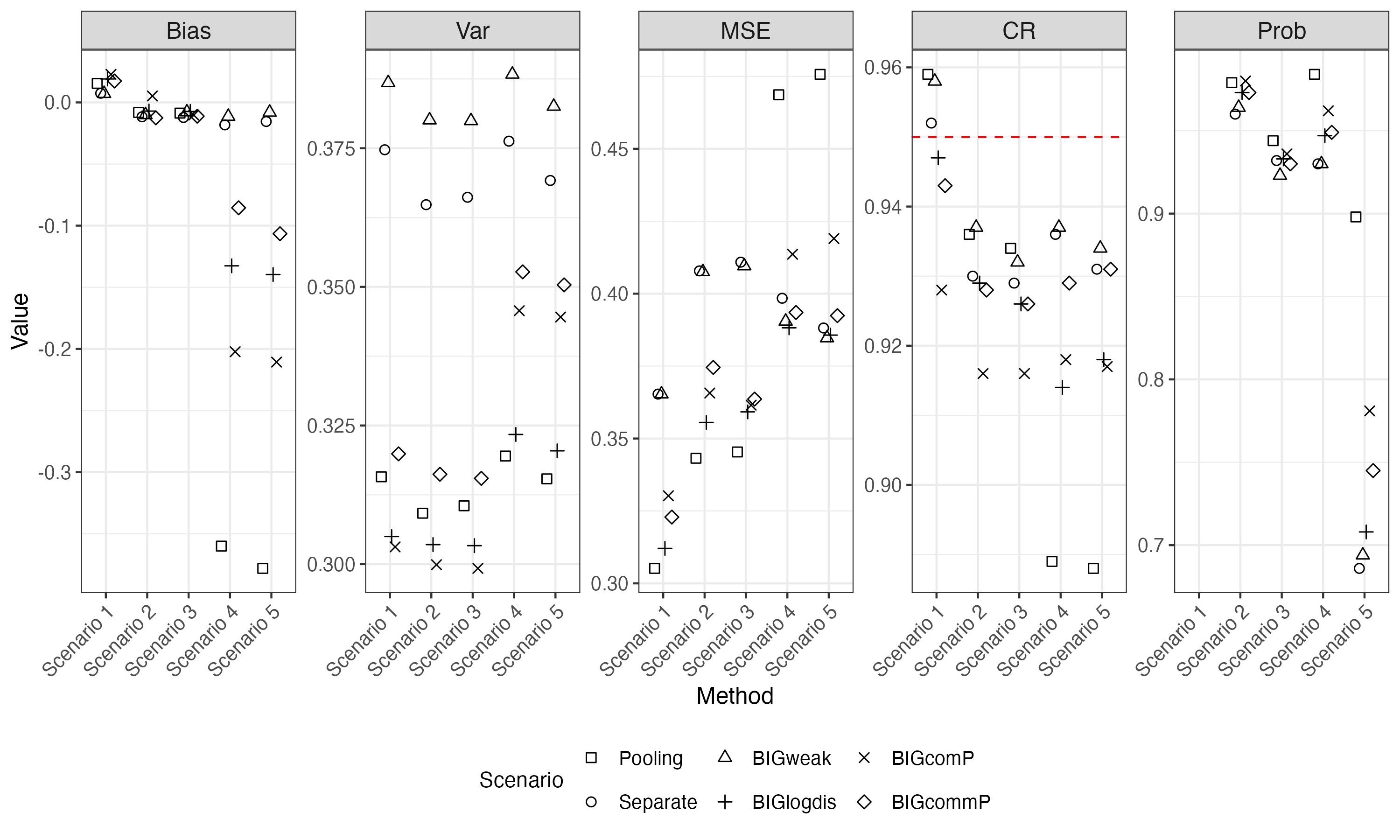"}
\caption{Simulation results with $n_{ori}=500$ and $r=0.3$. `Bias', `Var', `MSE', and `CR' represent the corresponding metrics (bias, variance, mean squared error, and cover rate) in terms of estimating $\mu_{11}-\mu_{31}$, i.e., the difference in expected outcome for DTRs $d_{11}$ and $d_{31}$ at cohort $c_2$. `Prob' represents the probability of identifying the true optimal DTR. BIGweak, BIGlogdis, BIGcomP and BIGcommP are the Bayesian integration g-formula (BIG) approaches with weakly informative priors, log distance priors, commensurate priors, and mixed commensurate priors.} \label{fig:sim1}
\end{figure}

\begin{figure}[H]
\centering
\includegraphics[width=0.8\textwidth]{"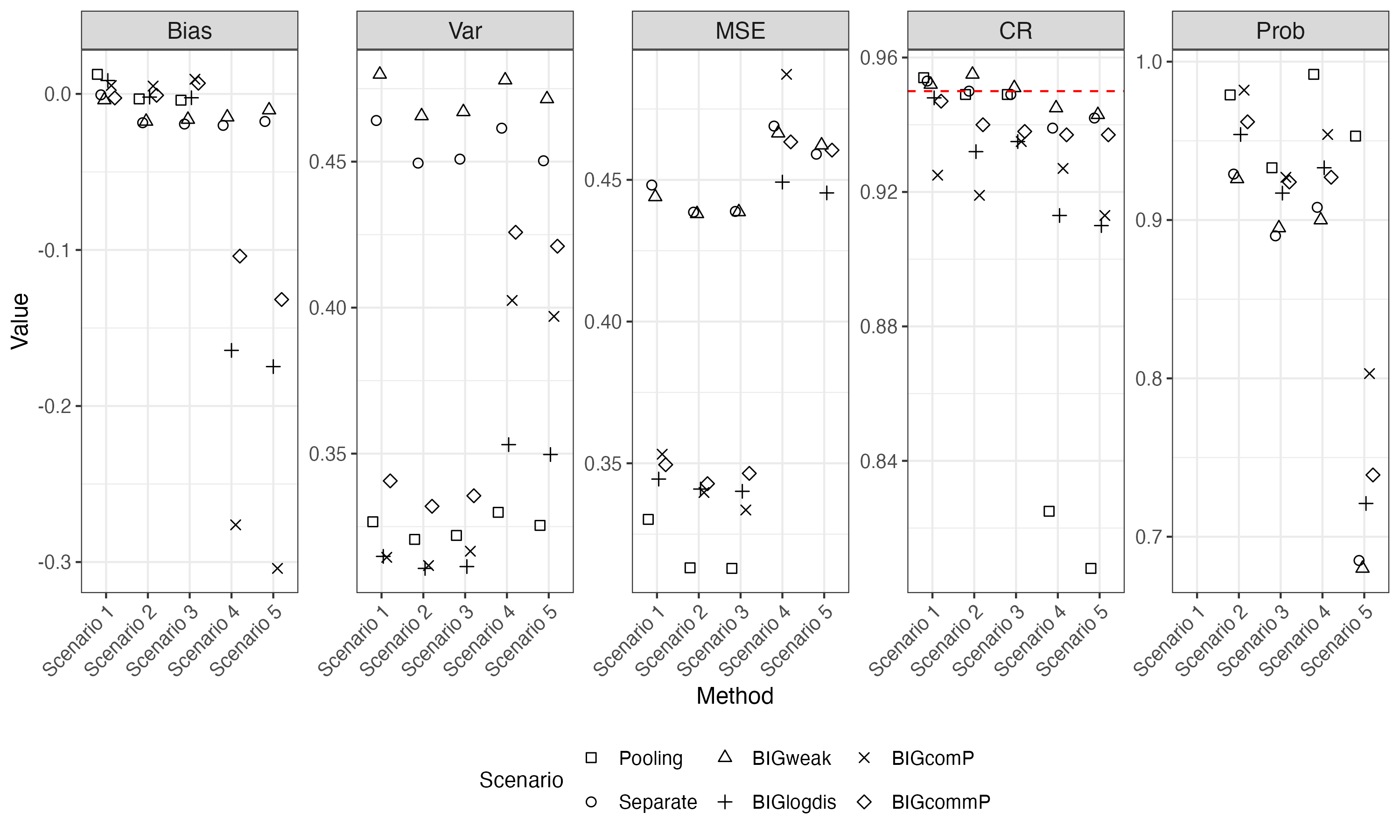"}
\caption{Simulation results with $n_{ori}=500$ and $r=0.5$. `Bias', `Var', `MSE', and `CR' represent the corresponding metrics (bias, variance, mean squared error, and cover rate) in terms of estimating $\mu_{11}-\mu_{31}$, i.e., the difference in expected outcome for DTRs $d_{11}$ and $d_{31}$ at cohort $c_2$. `Prob' represents the probability of identifying the true optimal DTR. BIGweak, BIGlogdis, BIGcomP and BIGcommP are the Bayesian integration g-formula (BIG) approaches with weakly informative priors, log distance priors, commensurate priors, and mixed commensurate priors.}
\end{figure}

\begin{figure}[H]
\centering
\includegraphics[width=0.8\textwidth]{"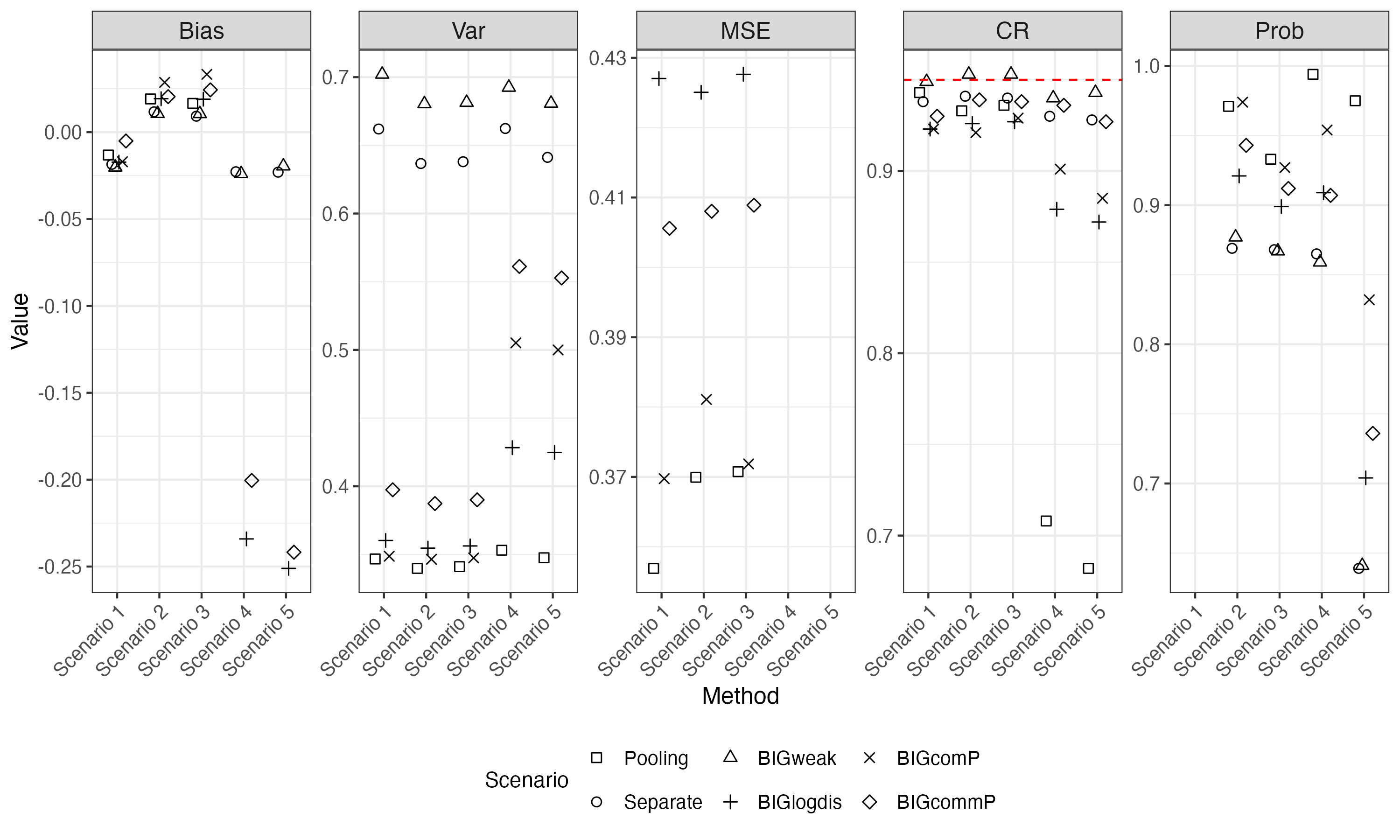"}
\caption{Simulation results with $n_{ori}=500$ and $r=0.7$. `Bias', `Var', `MSE', and `CR' represent the corresponding metrics (bias, variance, mean squared error, and cover rate) in terms of estimating $\mu_{11}-\mu_{31}$, i.e., the difference in expected outcome for DTRs $d_{11}$ and $d_{31}$ at cohort $c_2$. `Prob' represents the probability of identifying the true optimal DTR. BIGweak, BIGlogdis, BIGcomP and BIGcommP are the Bayesian integration g-formula (BIG) approaches with weakly informative priors, log distance priors, commensurate priors, and mixed commensurate priors.}
\end{figure}

\begin{figure}[H]
\centering
\includegraphics[width=0.8\textwidth]{"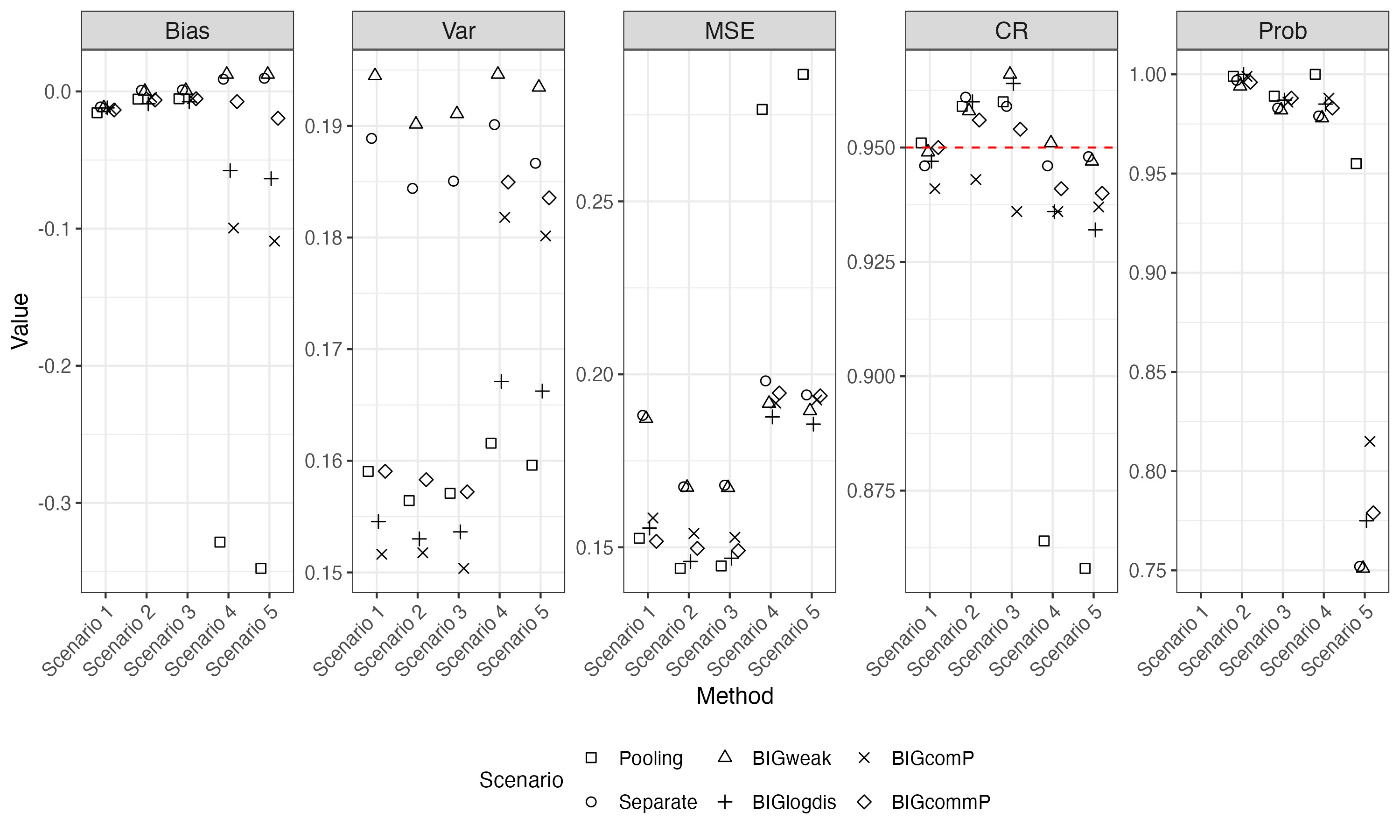"}
\caption{Simulation results with $n_{ori}=1000$ and $r=0.3$. `Bias', `Var', `MSE', and `CR' represent the corresponding metrics (bias, variance, mean squared error, and cover rate) in terms of estimating $\mu_{11}-\mu_{31}$, i.e., the difference in expected outcome for DTRs $d_{11}$ and $d_{31}$ at cohort $c_2$. `Prob' represents the probability of identifying the true optimal DTR. BIGweak, BIGlogdis, BIGcomP and BIGcommP are the Bayesian integration g-formula (BIG) approaches with weakly informative priors, log distance priors, commensurate priors, and mixed commensurate priors.}
\end{figure}

\begin{figure}[H]
\centering
\includegraphics[width=0.8\textwidth]{"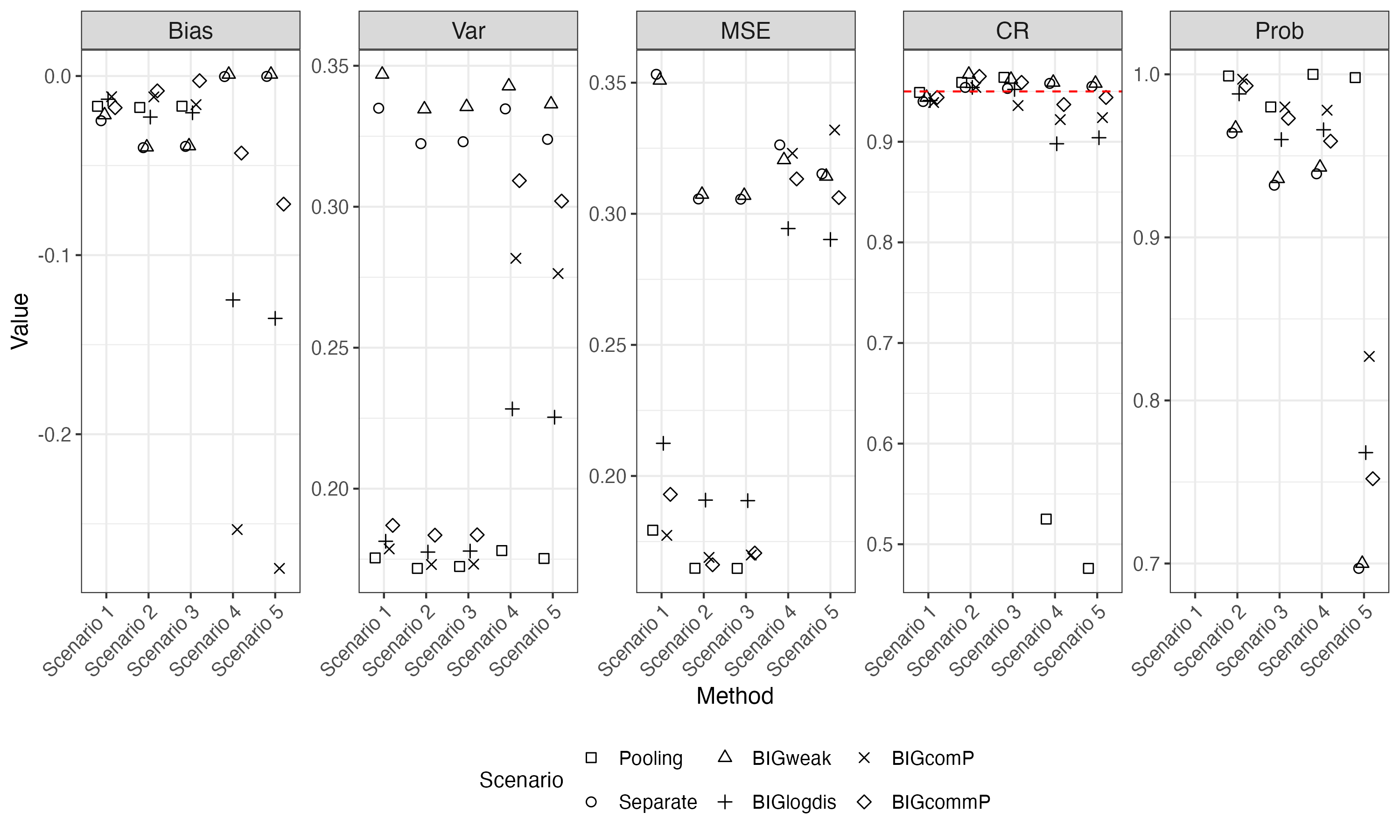"}
\caption{Simulation results with $n_{ori}=1000$ and $r=0.7$. `Bias', `Var', `MSE', and `CR' represent the corresponding metrics (bias, variance, mean squared error, and cover rate) in terms of estimating $\mu_{11}-\mu_{31}$, i.e., the difference in expected outcome for DTRs $d_{11}$ and $d_{31}$ at cohort $c_2$. `Prob' represents the probability of identifying the true optimal DTR. BIGweak, BIGlogdis, BIGcomP and BIGcommP are the Bayesian integration g-formula (BIG) approaches with weakly informative priors, log distance priors, commensurate priors, and mixed commensurate priors.}
\end{figure}

\begin{figure}[H]
\centering
\includegraphics[width=0.8\textwidth]{"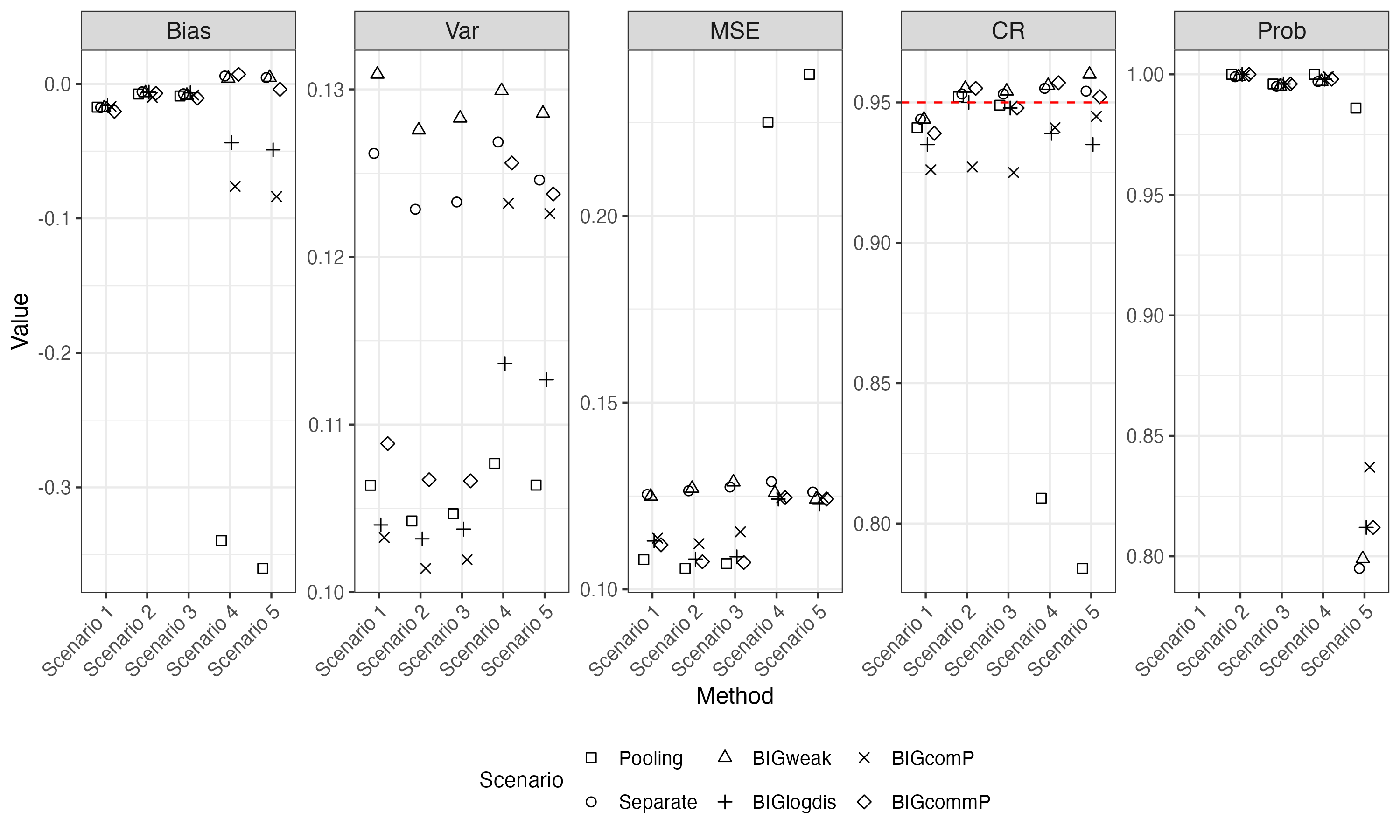"}
\caption{Simulation results with $n_{ori}=1000$ and $r=0.3$. `Bias', `Var', `MSE', and `CR' represent the corresponding metrics (bias, variance, mean squared error, and cover rate) in terms of estimating $\mu_{11}-\mu_{31}$, i.e., the difference in expected outcome for DTRs $d_{11}$ and $d_{31}$ at cohort $c_2$. `Prob' represents the probability of identifying the true optimal DTR. BIGweak, BIGlogdis, BIGcomP and BIGcommP are the Bayesian integration g-formula (BIG) approaches with weakly informative priors, log distance priors, commensurate priors, and mixed commensurate priors.}
\end{figure}

\begin{figure}[H]
\centering
\includegraphics[width=0.8\textwidth]{"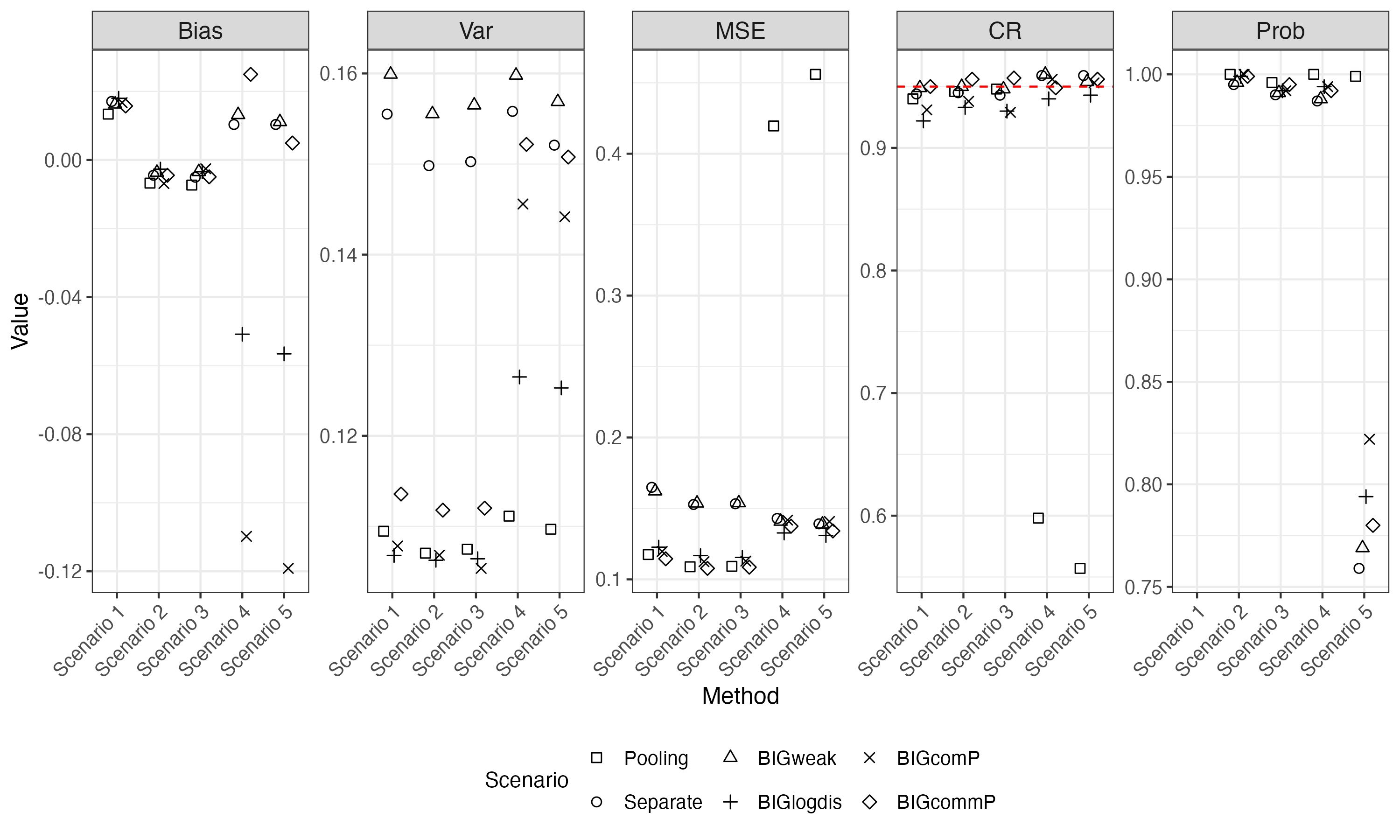"}
\caption{Simulation results with $n_{ori}=1500$ and $r=0.5$. `Bias', `Var', `MSE', and `CR' represent the corresponding metrics (bias, variance, mean squared error, and cover rate) in terms of estimating $\mu_{11}-\mu_{31}$, i.e., the difference in expected outcome for DTRs $d_{11}$ and $d_{31}$ at cohort $c_2$. `Prob' represents the probability of identifying the true optimal DTR. BIGweak, BIGlogdis, BIGcomP and BIGcommP are the Bayesian integration g-formula (BIG) approaches with weakly informative priors, log distance priors, commensurate priors, and mixed commensurate priors.}
\end{figure}

\begin{figure}[H]
\centering
\includegraphics[width=0.8\textwidth]{"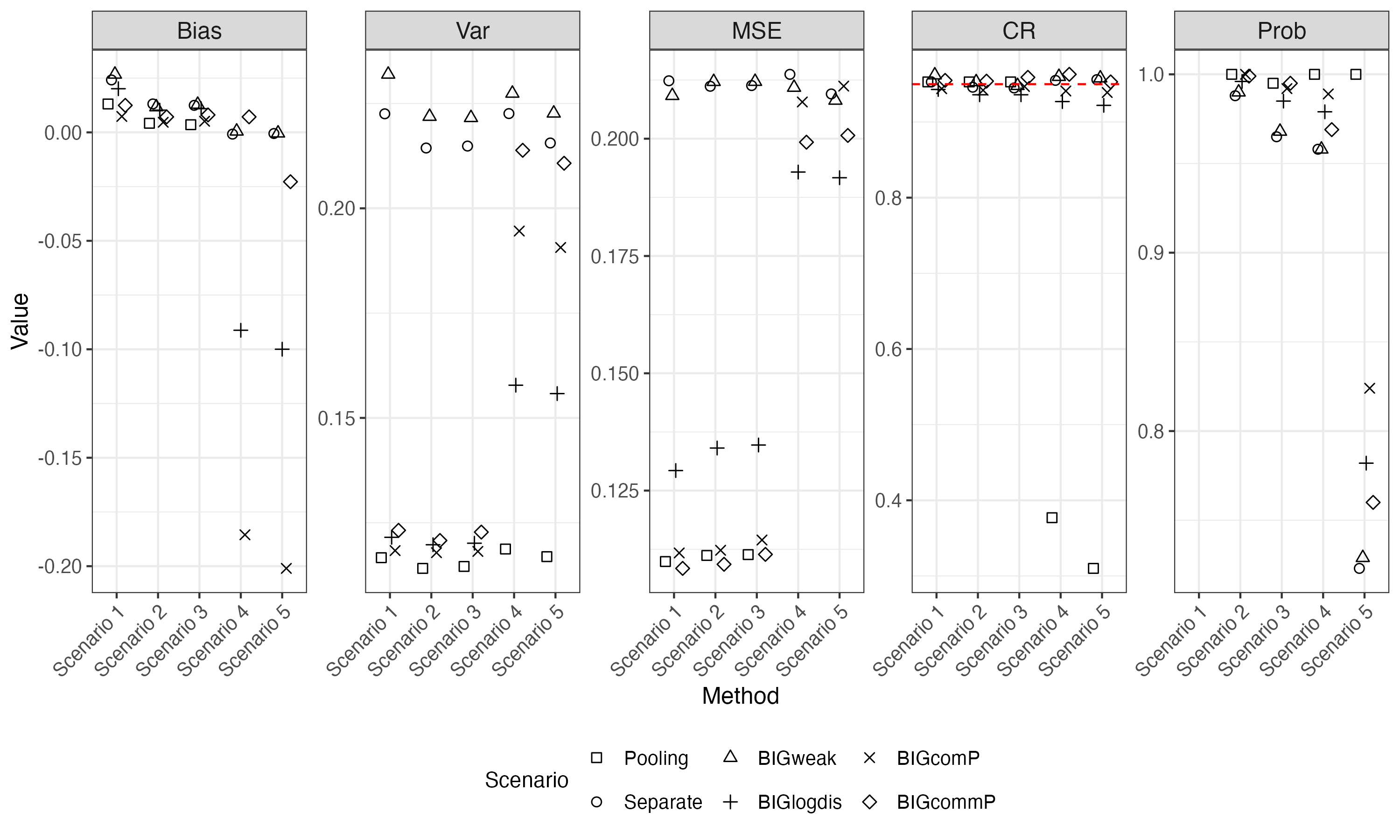"}
\caption{Simulation results with $n_{ori}=1500$ and $r=0.7$. `Bias', `Var', `MSE', and `CR' represent the corresponding metrics (bias, variance, mean squared error, and cover rate) in terms of estimating $\mu_{11}-\mu_{31}$, i.e., the difference in expected outcome for DTRs $d_{11}$ and $d_{31}$ at cohort $c_2$. `Prob' represents the probability of identifying the true optimal DTR. BIGweak, BIGlogdis, BIGcomP and BIGcommP are the Bayesian integration g-formula (BIG) approaches with weakly informative priors, log distance priors, commensurate priors, and mixed commensurate priors.}\label{fig:sim8}
\end{figure}

\section{Application results}

\begin{figure}[H]
\centering
\includegraphics[width=0.8\textwidth]{"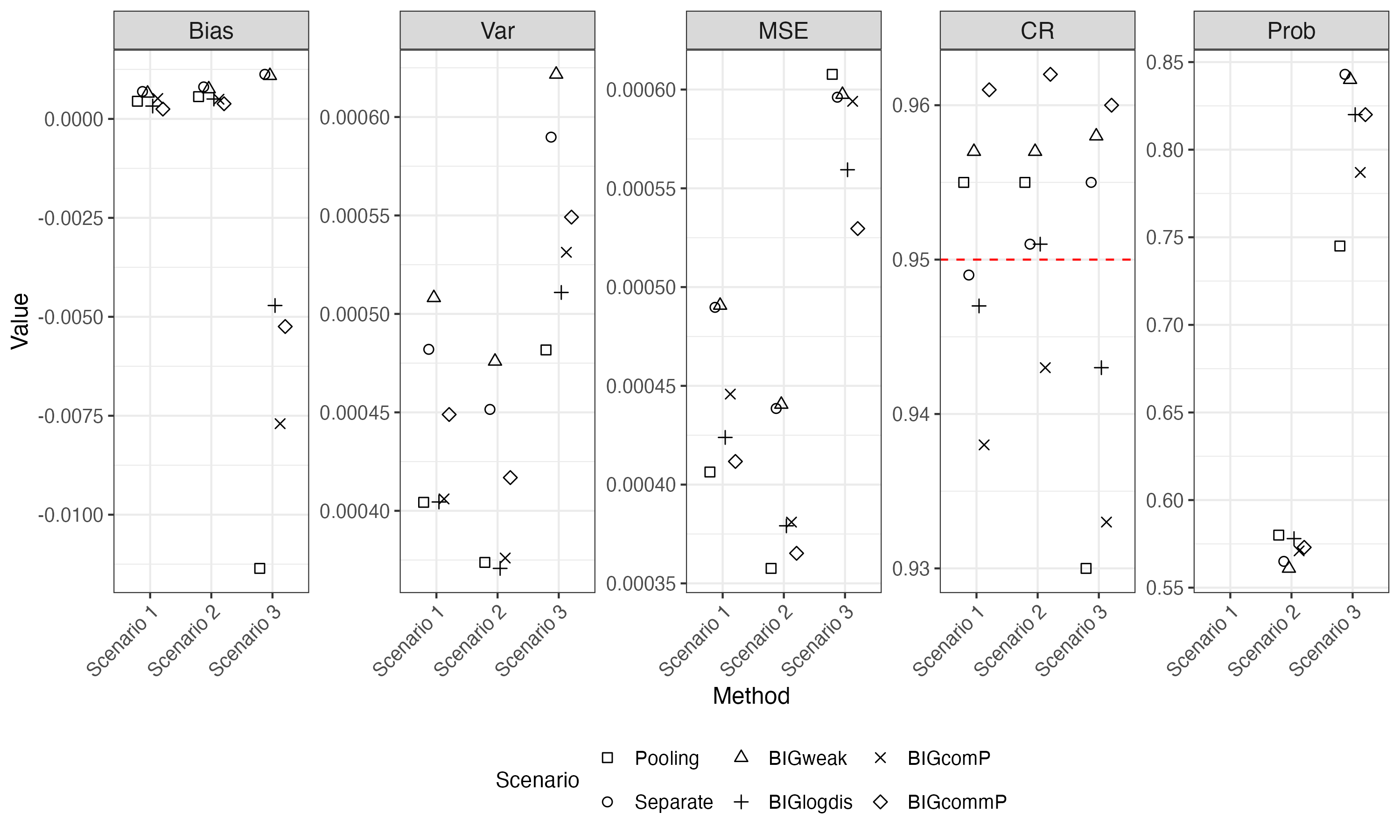"}
\caption{Application results with $n=2000$ and $r=0.3$. `Bias', `Var', `MSE', and `CR' represent the corresponding metrics (bias, variance, mean squared error, and cover rate) in terms of estimating $\mu_{11}-\mu_{31}$, i.e., the difference in expected outcome for DTRs $d_{11}$ and $d_{31}$ at cohort $c_2$. `Prob' represents the probability of identifying the true optimal DTR. BIGweak, BIGlogdis, BIGcomP and BIGcommP are the Bayesian integration g-formula (BIG) approaches with weakly informative priors, log distance priors, commensurate priors, and mixed commensurate priors.} \label{fig:app1}
\end{figure}

\begin{figure}[H]
\centering
\includegraphics[width=0.8\textwidth]{"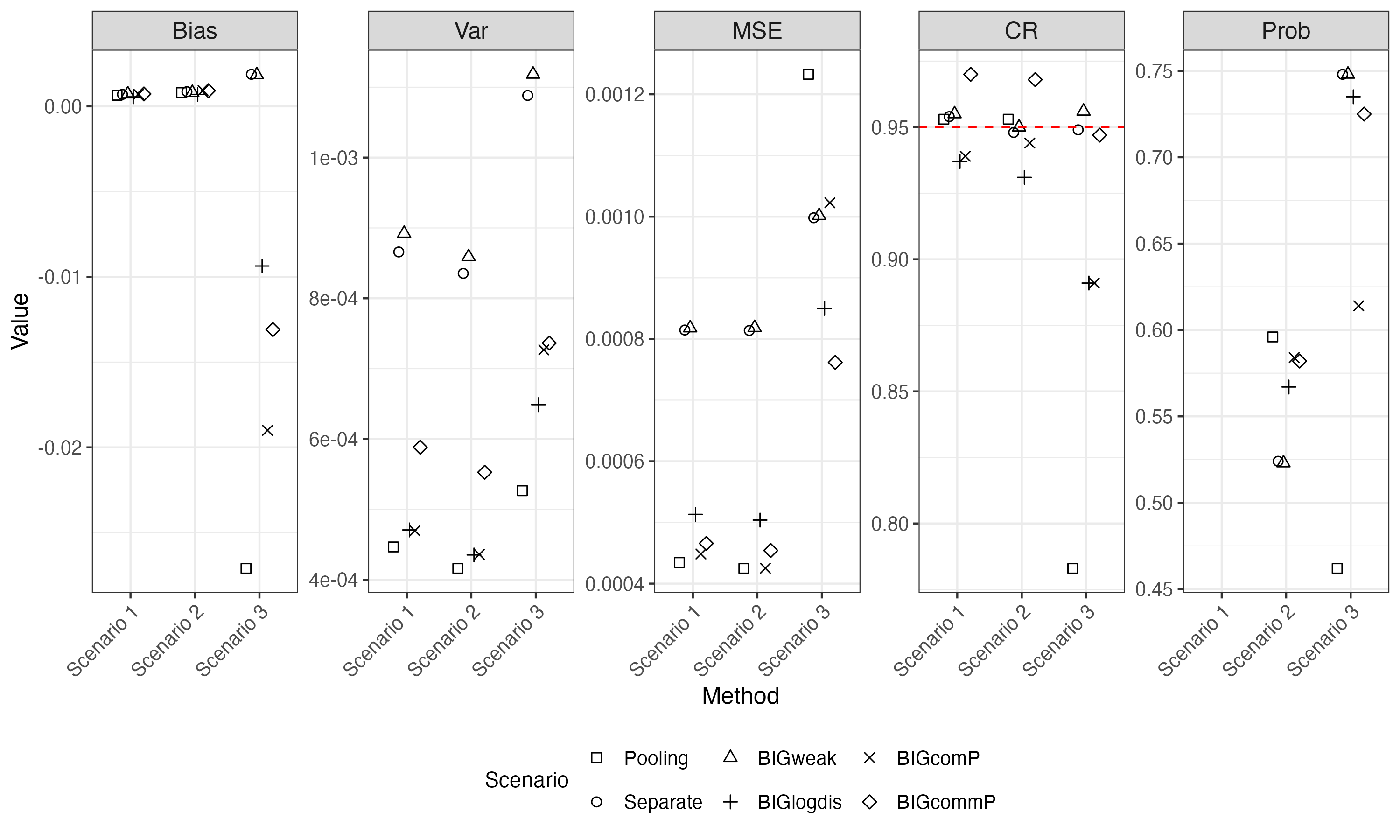"}
\caption{Application results with $n=2000$ and $r=0.7$. `Bias', `Var', `MSE', and `CR' represent the corresponding metrics (bias, variance, mean squared error, and cover rate) in terms of estimating $\mu_{11}-\mu_{31}$, i.e., the difference in expected outcome for DTRs $d_{11}$ and $d_{31}$ at cohort $c_2$. `Prob' represents the probability of identifying the true optimal DTR. BIGweak, BIGlogdis, BIGcomP and BIGcommP are the Bayesian integration g-formula (BIG) approaches with weakly informative priors, log distance priors, commensurate priors, and mixed commensurate priors.}
\end{figure}

\begin{figure}[H]
\centering
\includegraphics[width=0.8\textwidth]{"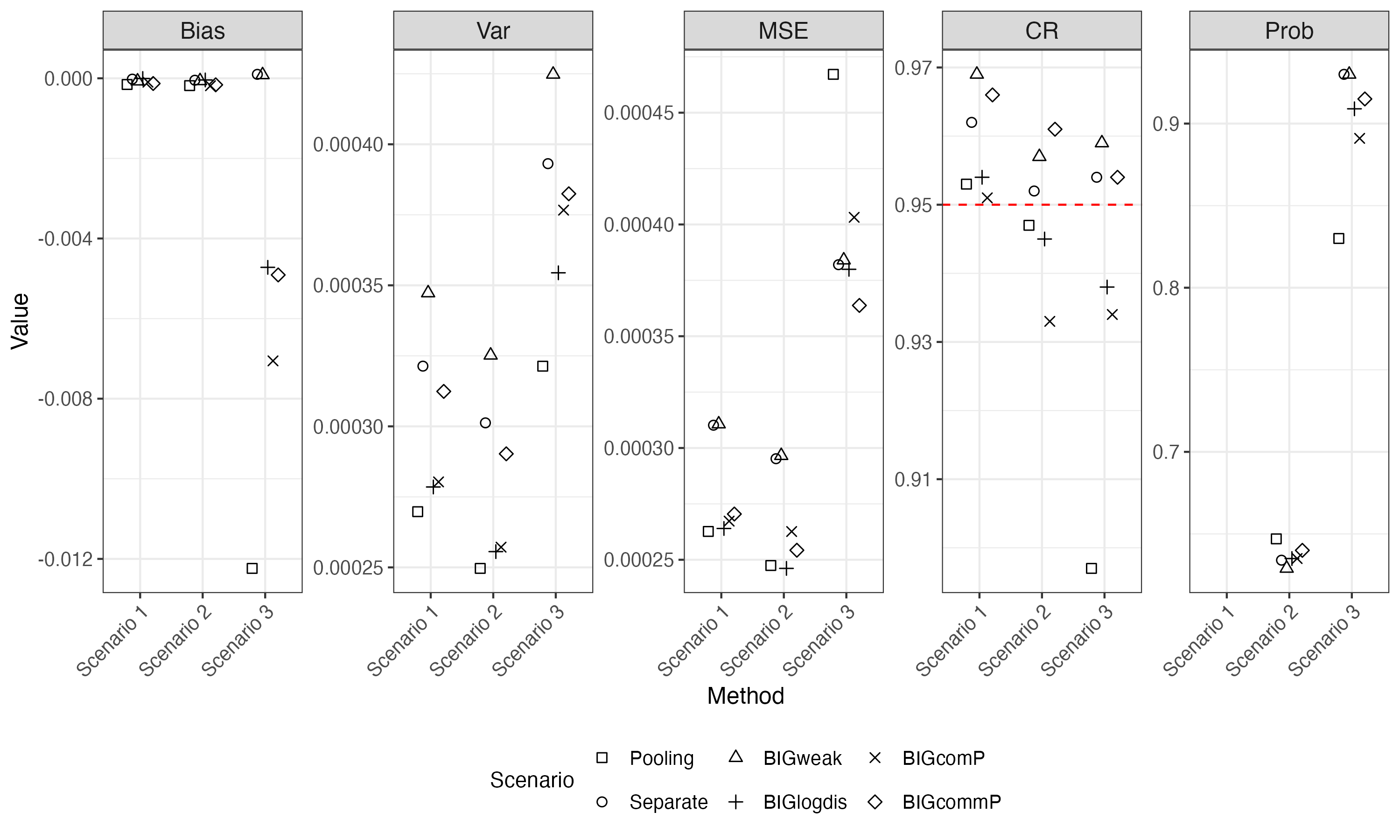"}
\caption{Application results with $n=3000$ and $r=0.3$. `Bias', `Var', `MSE', and `CR' represent the corresponding metrics (bias, variance, mean squared error, and cover rate) in terms of estimating $\mu_{11}-\mu_{31}$, i.e., the difference in expected outcome for DTRs $d_{11}$ and $d_{31}$ at cohort $c_2$. `Prob' represents the probability of identifying the true optimal DTR. BIGweak, BIGlogdis, BIGcomP and BIGcommP are the Bayesian integration g-formula (BIG) approaches with weakly informative priors, log distance priors, commensurate priors, and mixed commensurate priors.}
\end{figure}

\begin{figure}[H]
\centering
\includegraphics[width=0.8\textwidth]{"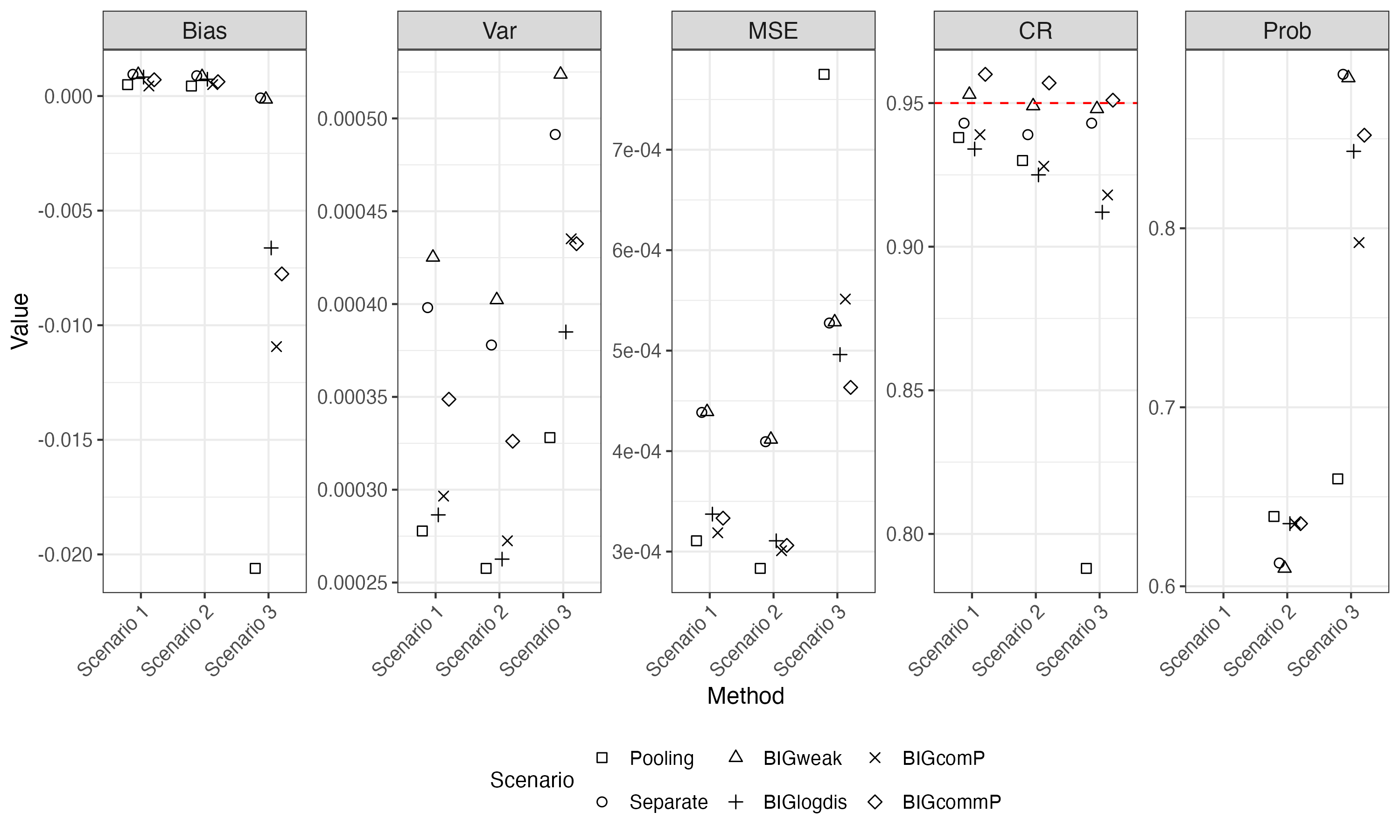"}
\caption{Application results with $n=3000$ and $r=0.5$. `Bias', `Var', `MSE', and `CR' represent the corresponding metrics (bias, variance, mean squared error, and cover rate) in terms of estimating $\mu_{11}-\mu_{31}$, i.e., the difference in expected outcome for DTRs $d_{11}$ and $d_{31}$ at cohort $c_2$. `Prob' represents the probability of identifying the true optimal DTR. BIGweak, BIGlogdis, BIGcomP and BIGcommP are the Bayesian integration g-formula (BIG) approaches with weakly informative priors, log distance priors, commensurate priors, and mixed commensurate priors.}
\end{figure}

\begin{figure}[H]
\centering
\includegraphics[width=0.8\textwidth]{"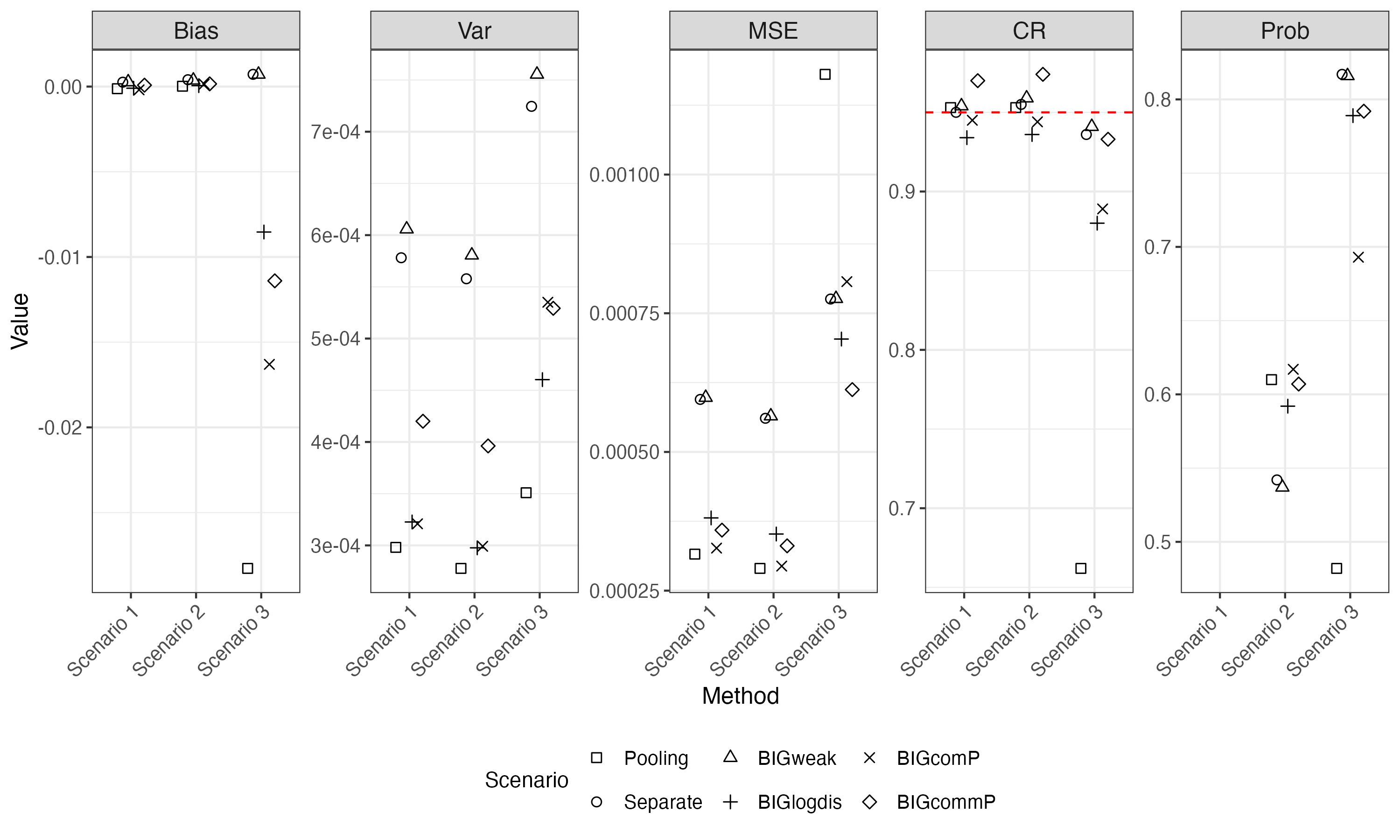"}
\caption{Application results with $n=3000$ and $r=0.7$. `Bias', `Var', `MSE', and `CR' represent the corresponding metrics (bias, variance, mean squared error, and cover rate) in terms of estimating $\mu_{11}-\mu_{31}$, i.e., the difference in expected outcome for DTRs $d_{11}$ and $d_{31}$ at cohort $c_2$. `Prob' represents the probability of identifying the true optimal DTR. BIGweak, BIGlogdis, BIGcomP and BIGcommP are the Bayesian integration g-formula (BIG) approaches with weakly informative priors, log distance priors, commensurate priors, and mixed commensurate priors.}
\end{figure}

\begin{figure}[H]
\centering
\includegraphics[width=0.8\textwidth]{"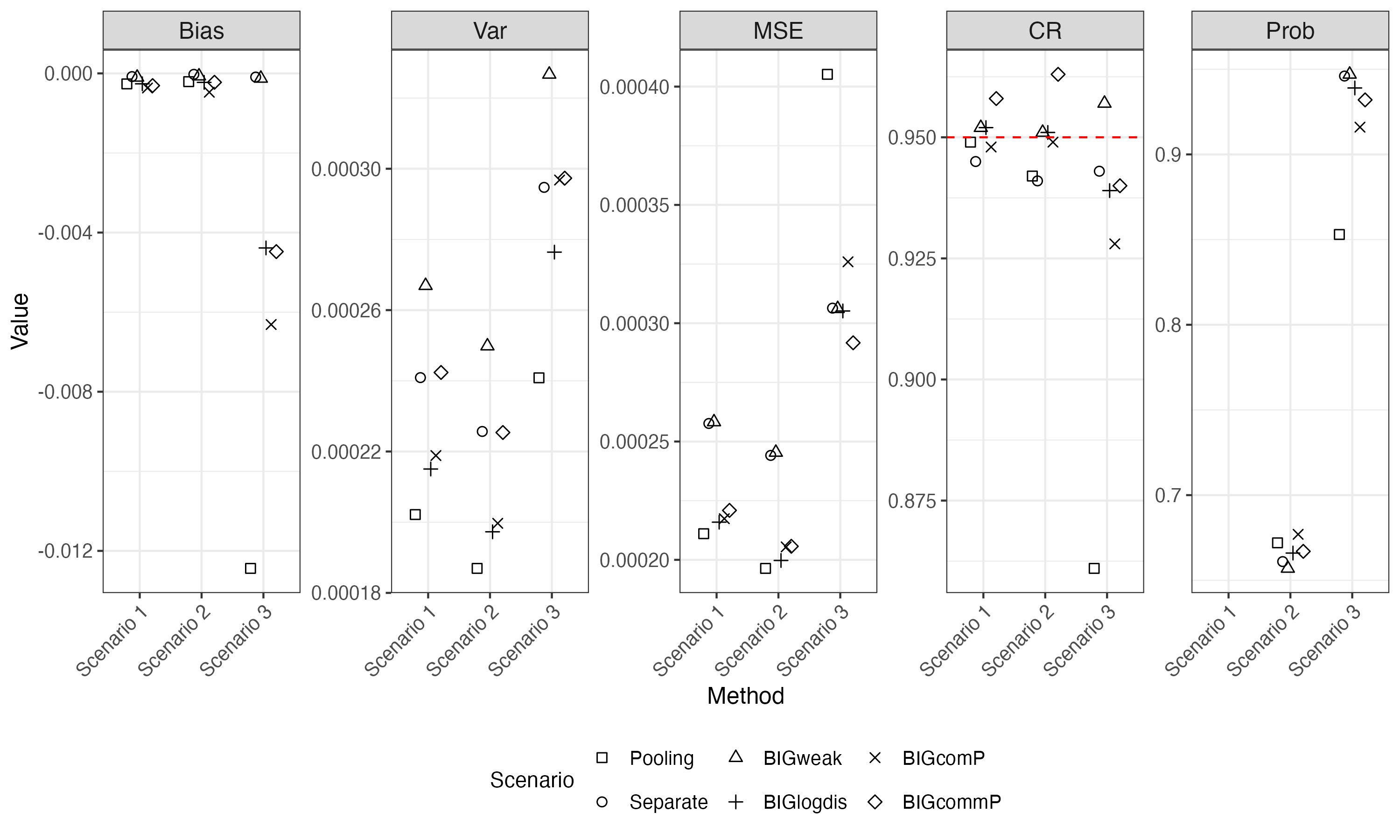"}
\caption{Application results with $n=4000$ and $r=0.3$. `Bias', `Var', `MSE', and `CR' represent the corresponding metrics (bias, variance, mean squared error, and cover rate) in terms of estimating $\mu_{11}-\mu_{31}$, i.e., the difference in expected outcome for DTRs $d_{11}$ and $d_{31}$ at cohort $c_2$. `Prob' represents the probability of identifying the true optimal DTR. BIGweak, BIGlogdis, BIGcomP and BIGcommP are the Bayesian integration g-formula (BIG) approaches with weakly informative priors, log distance priors, commensurate priors, and mixed commensurate priors.}
\end{figure}

\begin{figure}[H]
\centering
\includegraphics[width=0.8\textwidth]{"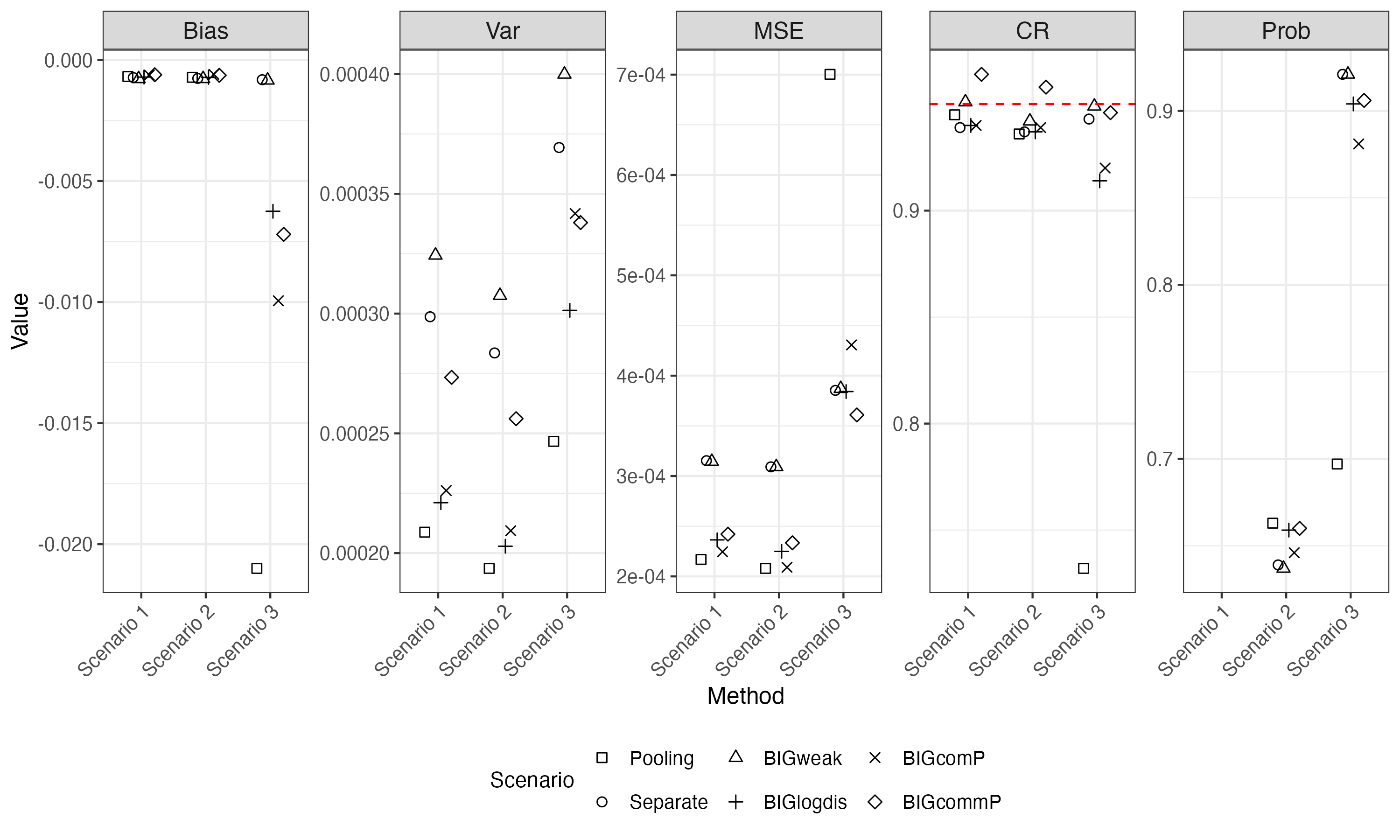"}
\caption{Application results with $n=4000$ and $r=0.5$. `Bias', `Var', `MSE', and `CR' represent the corresponding metrics (bias, variance, mean squared error, and cover rate) in terms of estimating $\mu_{11}-\mu_{31}$, i.e., the difference in expected outcome for DTRs $d_{11}$ and $d_{31}$ at cohort $c_2$. `Prob' represents the probability of identifying the true optimal DTR. BIGweak, BIGlogdis, BIGcomP and BIGcommP are the Bayesian integration g-formula (BIG) approaches with weakly informative priors, log distance priors, commensurate priors, and mixed commensurate priors.}
\end{figure}

\begin{figure}[H]
\centering
\includegraphics[width=0.8\textwidth]{"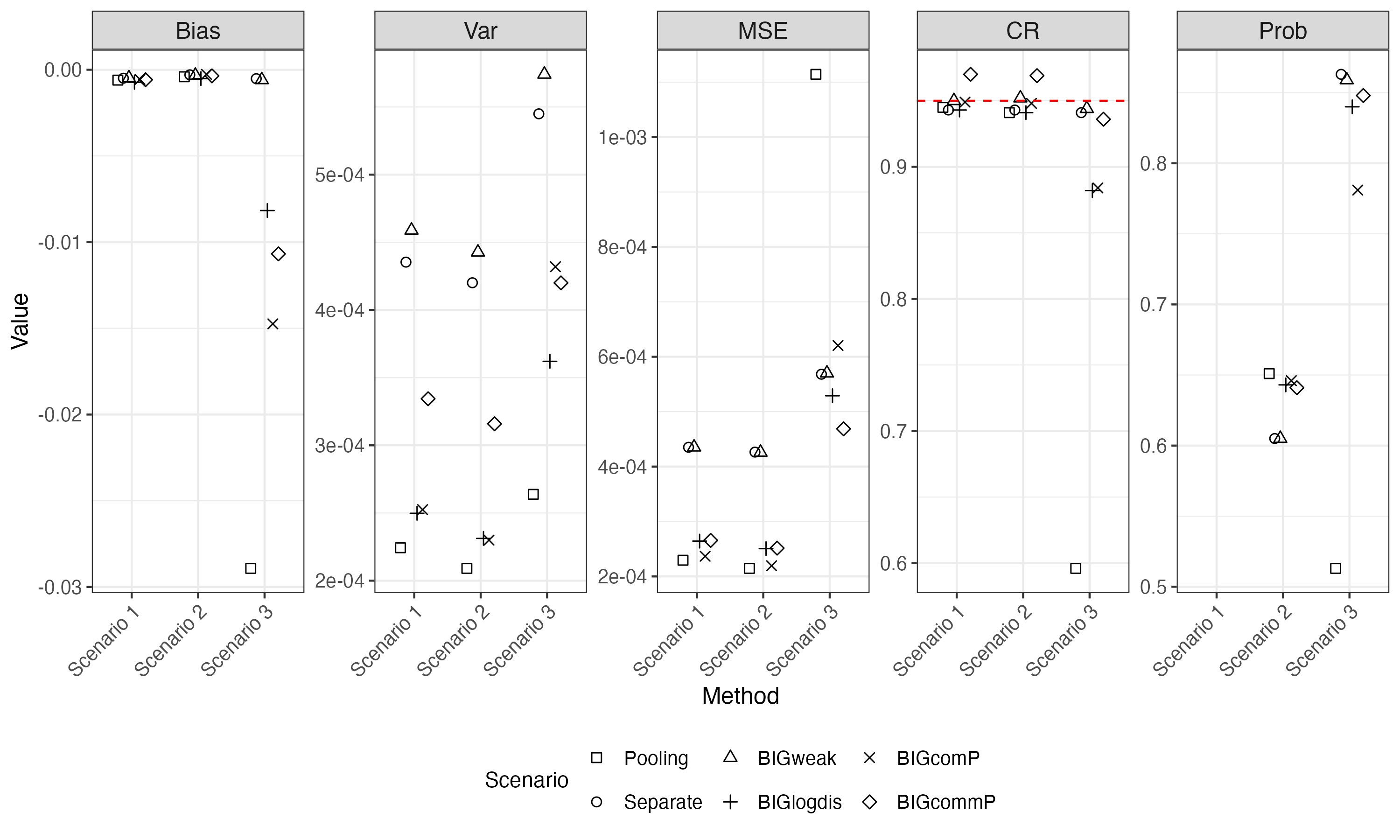"}
\caption{Application results with $n=4000$ and $r=0.7$. `Bias', `Var', `MSE', and `CR' represent the corresponding metrics (bias, variance, mean squared error, and cover rate) in terms of estimating $\mu_{11}-\mu_{31}$, i.e., the difference in expected outcome for DTRs $d_{11}$ and $d_{31}$ at cohort $c_2$. `Prob' represents the probability of identifying the true optimal DTR. BIGweak, BIGlogdis, BIGcomP and BIGcommP are the Bayesian integration g-formula (BIG) approaches with weakly informative priors, log distance priors, commensurate priors, and mixed commensurate priors.} \label{fig:app8}
\end{figure}

\end{document}